\documentclass[twocolumn, twocolappendix,]{aastex631}
\usepackage{natbib}
\usepackage{graphicx}
\usepackage{CJK}
\usepackage[version=4]{mhchem}
\usepackage{amsmath}
\usepackage{subfigure}
\usepackage{multirow}
\usepackage{wrapfig}
\usepackage{color,xcolor}
\usepackage{xspace}
\usepackage{float}

\hypersetup{linkcolor=magenta, citecolor=teal, urlcolor=teal}

\graphicspath{{./}{figures/}}

\newcommand{\Ha}{\ifmmode {\rm H}\alpha \else H$\alpha$\fi\xspace}
\newcommand{\Hb}{\ifmmode {\rm H}\beta \else H$\beta$\fi\xspace}
\newcommand{\hi}{\ifmmode \rm{H}\,\textsc{i} \else H\,{\sc i}\fi\xspace}

\newcommand{\Hii}{\ifmmode \rm{H}\,\textsc{ii} \else H\,{\sc ii}\fi}

\newcommand{\nii}{\ifmmode [\rm{N}\,\textsc{ii}] \else [N\,{\sc ii}]\fi\xspace}

\newcommand{\oi}{\ifmmode [\rm{O}\,\textsc{i}] \else [O\,{\sc i}]\fi\xspace}

\newcommand{\neiii}{\ifmmode [\rm{Ne}\,\textsc{iii}] \else [Ne\,{\sc iii}]\fi}

\newcommand{\hei}{\ifmmode \rm{He}\,\textsc{i} \else He\,{\sc i}\fi\xspace}
\newcommand{\heii}{\ifmmode \rm{He}\,\textsc{ii} \else He\,{\sc ii}\fi\xspace}
\newcommand{\oii}{\ifmmode [\rm{O}\,\textsc{ii}] \else [O\,{\sc ii}]\fi\xspace}

\newcommand{\oiii}{\ifmmode [\rm{O}\,\textsc{iii}] \else [O\,{\sc iii}]\fi\xspace}

\newcommand{\sii}{\ifmmode [\rm{S}\,\textsc{ii}] \else [S\,{\sc ii}]\fi\xspace}
\newcommand{\siii}{\ifmmode [\rm{S}\,\textsc{iii}] \else [S\,{\sc iii}]\fi}

\received{2021.10.28}
\revised{2021.12.13}
\accepted{2021.12.14}
\submitjournal{ApJ}

\shorttitle{Compactness of Galaxy Groups}
\shortauthors{Zheng et al.}

\begin{document}
\begin{CJK*}{UTF8}{gbsn}

\title{The Compactness of Galaxy Groups in the Sloan Digital Sky Survey}

\correspondingauthor{Shi-Yin Shen}
\email{ssy@shao.ac.cn}

\author[0000-0002-5632-9345]{Yun-Liang Zheng (郑云亮)}
\affiliation{Key Laboratory for Research in Galaxies and Cosmology, Shanghai Astronomical Observatory, Chinese Academy of Sciences,  80 Nandan Road, Shanghai, China, 200030}
\affiliation{University of Chinese Academy of Sciences,  No.19A Yuquan Road, Beijing, China, 100049}

\author[0000-0002-3073-5871]{Shi-Yin Shen (沈世银)}
\affiliation{Key Laboratory for Research in Galaxies and Cosmology, Shanghai Astronomical Observatory, Chinese Academy of Sciences,  80 Nandan Road, Shanghai, China, 200030}
\affiliation{Key Lab for Astrophysics, Shanghai, China, 200034}

\author[0000-0002-9767-9237]{Shuai Feng (冯帅)}
\affiliation{College of Physics, Hebei Normal University, 20 South Erhuan Road, Shijiazhuang, 050024, China}
\affiliation{Hebei Key Laboratory of Photophysics Research and Application, 050024 Shijiazhuang, China}

\begin{abstract}
We use an updated version of the halo-based galaxy group catalog of Yang et al., and take the surface brightness of the galaxy group ($\mu_{\rm lim}$) based on projected positions and luminosities of galaxy members as a compactness proxy to divide groups into sub-systems with different compactness. By comparing various properties, including galaxy conditional luminosity function, stellar population, active galactic nuclei (AGN) activity, and X-ray luminosity of the intra-cluster medium of carefully controlled high (HC) and low compactness (LC) group samples, we find that the group compactness plays an essential role in characterizing the detailed physical properties of the group themselves and their group members, especially for low mass groups with $M_h \lesssim 10^{13.5}h^{-1}M_{\odot}$. We find that the low-mass HC groups have a systematically lower magnitude gap $\Delta m_{12}$ and X-ray luminosity than their LC counterparts, indicating that the HC groups are probably in the early stage of group merging. On the other hand, a higher fraction of passive galaxies is found in the HC group, which however is a result of systematically smaller halo-centric distance distribution of their satellite population. After controlling of both $M_h$ and halo-centric distance, we do not find any differences for both the quenching faction and AGN activity of the member galaxies between the HC and LC groups. Therefore, we conclude that the halo quenching effect, which result in the halo-centric dependence of galaxy population, is a faster process compared to the dynamical relaxed time-scale of galaxy groups.
\end{abstract}


\section{Introduction} \label{sec:intro}
In the paradigm of hierarchical structure formation, galaxies are believed form via baryonic gas cooling and condensation within dark matter halos. It is generally assumed that halo mass ($M_h$) is the most fundamental parameter that determines the evolutionary path of  galaxies inside \citep[e.g.,][]{Weinmann..2006, Wetzel..2012, Woo..2013,  Woo..2015, Bluck..2014, Bluck..2016, WangEnci..2018, Davies..2019}. Based on this assumption, numerous models for linking galaxies to dark matter halos have been developed, such as the halo occupation distribution model \citep[HOD,][]{JingYP..1998, Peacock.Smith2000, ZhengZ..2005, Zu.Mandelbaum2016}, abundance matching model \citep[HAM,][]{MoHJ..1999, Kravtsov..2004, Vale.Ostriker2006, GuoQ..2010, Simha..2012, Reddick..2013}, the conditional luminosity function \citep[CLF,][]{Yang..2003, Yang..2008, Cooray2006, vandenBosch..2007}, and halo-based empirical models \citep[][]{Yang..2013, LuZK..2014, Moster..2018}. In these models, dark matter halos have evolved hierarchically through repeated mergers. Galaxies that form the main branch of a merger tree generally reside near the halo center \citep{Lange..2018} and are referred to as central galaxies, while others that form in the sub-branch and orbit the central galaxies are referred to as satellite galaxies. These two galaxy populations have experienced different evolutionary paths so that their properties are different statistically as a result \citep[e.g.,][]{Weinmann..2006, vandenBosch..2008, PengYJ..2010, Knobel..2015, Davies..2019}.

From an observational point of view, the physical properties of galaxies are strongly correlated with their stellar mass ($M_{\star}$) and environment \citep[e.g.,][]{Dekel.Birnboim2006, PengYJ..2010, PengYJ..2012, PengYJ..2014, LiuCX..2019, ZhangCP..2021}. Specific mechanisms have been proposed for mass dependency, including supernovae and active galactic nuclei (AGN) feedback \citep[][]{Croton..2006, Fabian2012, Harrison2017}, while the physical mechanisms of the environmental effects on the dwarf/satellite galaxies include tidal interaction \citep{Farouki.Shapiro1981}, ram-pressure stripping \citep{Gunn.Gott1972}, and harassment \citep{Moore..1996}, etc. In the framework of halo model, the observed environmental effects of galaxies can be interpreted as a dependence on their host halos \citep[e.g.,][]{LiuL..2010, Wetzel..2012, Henriques..2017, WangEnci..2018, Davies..2019}.  Indeed, \citet{Haas..2012} demonstrate that the environmental parameters of galaxies defined in a traditional way (e.g., projected local density, projected distance to the $i$th nearest galaxy, etc.) are in effect measures of their host halo mass $M_h$. However, there are also observational indications that galaxies exhibit more complex environmental effects beyond the simple $M_h$ model. For example, \citet{More..2011} and \citet{Mandelbaum..2016} revealed that the stellar-to-halo mass relation is further correlated with the physical properties of central galaxies. From an evolutionary point of view, dark matter halos form via mergers of smaller halos. When dark matter halos merge, infalling galaxies do not necessarily merge with central galaxies in a short timescale but keep orbiting and interacting with the central galaxies \citep{Stewart..2009}. Therefore, in addition to $M_h$, the halo formation time (or the assembly history) has long been used as the second parameter to profile the dark matter halos when analyzing the simulation data \citep[e.g.,][]{Matthee..2017, Zehavi..2018, CuiWG..2021}.

In observation, the host $M_h$ of galaxy systems can be estimated from the total stellar mass or richness of the group members \citep[e.g.,][]{Yang..2005, Yang..2007, Yang..2021, Robotham..2011}. However, it is not clear how to link the observational properties of galaxy groups to their assembly history. Several studies on fossil groups of galaxies \citep[e.g.,][]{Dariush..2010, Raouf..2014, Raouf..2018, Khosroshahi..2017, Li.Cen2020} have suggested that $\Delta m_{12}$, the magnitude gap between the first and second brightest members, might characterize the assembly history of a galaxy group. However, recent studies \citep[e.g.,][]{Shen..2014, Trevisan.Mamon2017} have shown that the origin of large $\Delta m_{12}$ is quite complex and cannot be simply accounted for by the halo formation time. On the other hand, like the size of galaxies, the size (or compactness) of galaxy groups might be another observationally accessible parameter encoding the assembly histories of halos. Indeed, it has long been known that there is a special class of galaxy groups, compact groups of galaxies (CGs), which is defined as a group system containing a few galaxies separated by distance of an order of galaxy size \citep[e.g.,][]{Hickson1982, Barton..1996, DiazGimenez.Zandivarez2015, Sohn..2015, Sohn..2016, DiazGimenez..2018, Paper1}. Because of the observational facts that tidal interactions and mergers frequently appear among CG members, \citep[e.g.,][]{Barnes1989, Amram..2004, Coziol.Plauchu-Frayn2007, Torres-Flores..2009, Torres-Flores..2010, Torres-Flores..2014}, CGs are believed to represent the sites for massive members that are about to merge. Moreover, based on simulation, \citet{Farhang..2017} suggested that CGs are early-formed systems that depart from ordinary hierarchical assembly and stressed that CGs appear to be a distinct halo environment. Inspired by the ideas about the peculiarity of CGs, several comparative studies on the galaxy properties of CGs and non-CGs have been performed. For instance, \citet{Coenda..2012} showed that CGs contain a larger fraction of red, early-type galaxies in comparison to loose groups and fields at a given absolute magnitude, while \citet{Cluver..2020} found no significant difference in the scale relation between quenched fraction and $M_{\star}$ in group systems with different levels of compactness. 

The connection between the observational selected CGs and halo-model-based groups has been recently investigated by \citet{Paper2}. In that study, we show that the observationally selected CGs are physically heterogeneous systems and can be mainly separated into two categories, the isolated systems and those embedded in rich groups or clusters. For these isolated CGs, we find that they have lower dynamical masses compared to the non-compact ones at the same group luminosity. However, we cannot distinguish whether this lower dynamical mass state is a result of a merging process or the dynamical evolution of post-mergers. In this paper, we expand our studies on CGs \citep{Paper1,Paper2} and for the first time, take the compactness of galaxy groups as a general parameter to explore its correlations with other physical properties of galaxy groups and group members when $M_h$ is controlled. Moreover, since the X-ray emission from the intragroup medium contains rich information on the dynamical status of galaxy groups, for the first time, we also compare the X-ray properties of galaxy groups with different compactness using a statistical method in this study. Our goal is to figure out whether the group compactness plays a role as a second observational parameter in describing the properties and merger histories of galaxy groups.

This paper is organized as follows. We describe the halo-based group samples in Section~\ref{sec:data}, and the measurements of relevant properties in section~\ref{sec:props}. We present our results for the global and individual properties of groups with different compactness in section~\ref{sec:global} and~\ref{sec:properties}, respectively. We provide a brief discussion in Section~\ref{sec:discussion} and summarize our conclusions in Section~\ref{sec:conclusion}. Throughout this paper, we assume a flat $\Lambda$CDM cosmology with parameters: $\Omega_{m} = 0.27$ and $H_0 = 100 h$ km s$^{-1}$ Mpc$^{-1}$ with $h = 0.7$.

\section{Sample: groups of galaxies} \label{sec:data}
In this work, we take use of the group catalog of \citet[][hereafter Y07]{Yang..2007, Yang..2012}, which has been derived from the latest version of New York University Value-Added Galaxy Catalog \citep[VAGC,][]{NYU-VAGC} based on the SDSS legacy survey \citep{SDSSDR7} with a set of improved reduction. Y07 has applied a halo-based group finder to assign each galaxy in the SDSS-DR7 Main Galaxy Sample (MGS) within the redshift range of $0.01 < z < 0.20$ and $r$-band magnitude brighter than $17.72$ mag to a unique group. In Y07, three versions of a group catalog have been constructed based on different redshift sources. In this work, we use sample III of the Y07 group catalog containing $\sim 640,000$ galaxies, where a small fraction of the members ($\sim 37,000$) without spectroscopic redshifts have been assigned with the redshifts of their nearest neighbors. Therefore, sample III of the Y07 group catalog has $100 \%$ completeness of member galaxies, while it also contains contamination from background or foreground galaxies.

Since $\sim 13,000$ out of $\sim 37,000$ spectroscopically unmeasured galaxies have achieved new redshifts from SDSS-DR16 \citep{SDSSDR16}, GAMA-DR2 \citep{gama}, and LAMOST-DR7 \citep{lamost}, the contamination of the group catalog can be significantly reduced \citep[e.g.,][]{Shen..2016, Feng..2019, Paper1}. Here, we use these new redshift measurements to update the sample III of the Y07 group catalog. We find $\sim 8500$ out of $\sim 13,000$ galaxies with new redshift measurements are concordant with their nearest neighbors, thus, their group ID remains unchanged. For the remaining $\sim 4500$ galaxies having updated redshifts but inconsistent with their nearest neighbors, we reassign them to a particular group following the criteria of Y07 and update the original halo properties accordingly. For the remaining $\sim 24,000$ out of $\sim 37,000$ galaxies without new redshift detection, we keep them unchanged. 

Owing to our aim of investigating how group properties vary with group compactness, which is defined based on at least three members, we only consider $N \ge 3$ groups in this work. Moreover, we have also noticed that bad photometry \citep[e.g., one large extended galaxy deblending into various small pieces or a fragment of a satellite trail identified as a chain of several galaxies,][]{McConnachie..2009, Paper1} will contaminate the following results because these problematic sources are more likely to be identified as groups with higher compactness. Therefore, we have made visual inspections of the images of the $N \ge 3$ groups and removed $\sim 300$ fake detections, most of which belong to low $M_h$ groups.

Y07 has estimated the characteristic stellar mass ($M_{19.5}$) and luminosity ($L_{19.5}$) of each group by summing up the masses and luminosities of all the members with $^{0.1}M_{r} - 5 \log h \le -19.5$ mag, respectively. For the distant groups ($z > 0.09$) where the corresponding faint-end limit brighter than $^{0.1}M_{r} - 5 \log h = -19.5$ mag, their $M_{19.5}$ and $L_{19.5}$ have been corrected for the missing members. Y07 has also estimated two versions of $M_{h}$ for each group based on their $M_{19.5}$ or $L_{19.5}$, respectively. Unless stated otherwise, we adopt the values based on $M_{19.5}$ as recommended in Y07 paper. For some groups, their $M_h$ have been updated due to the slight revision of their group membership, especially for those groups containing the galaxies with discordant new redshifts. There are also $\sim 700$ $N \ge 3$ groups without $M_h$ calculated because none of their members is brighter than $^{0.1}M_{r} - 5 \log h \le -19.5$ mag, thus we remove them in the following analysis. 

Finally, there are 22,468 $N \ge 3$ groups with $M_h$ assigned, which have 140,930 member galaxies and 129,622 of them ($\sim 92 \%$) have spectroscopic redshifts. The lower redshift completeness of the final sample compared to the overall galaxy sample ($\sim 95 \%$) is caused by the fact that almost all of the galaxies with assigned redshifts are members of $N \ge 2$ systems.

In Y07, the dark matter halos are defined as having an over-density of 180, we recompute the halo radius ($R_{180}$) and group velocity dispersion ($\sigma$) given by the Equation (5) and (6) in Y07 paper due to the revision of $M_h$ estimate. Unless stated otherwise, the central galaxy of each group are identified to be the most massive one.

\begin{figure*}
\centering
  \subfigure{
    \includegraphics[width=1.\hsize]{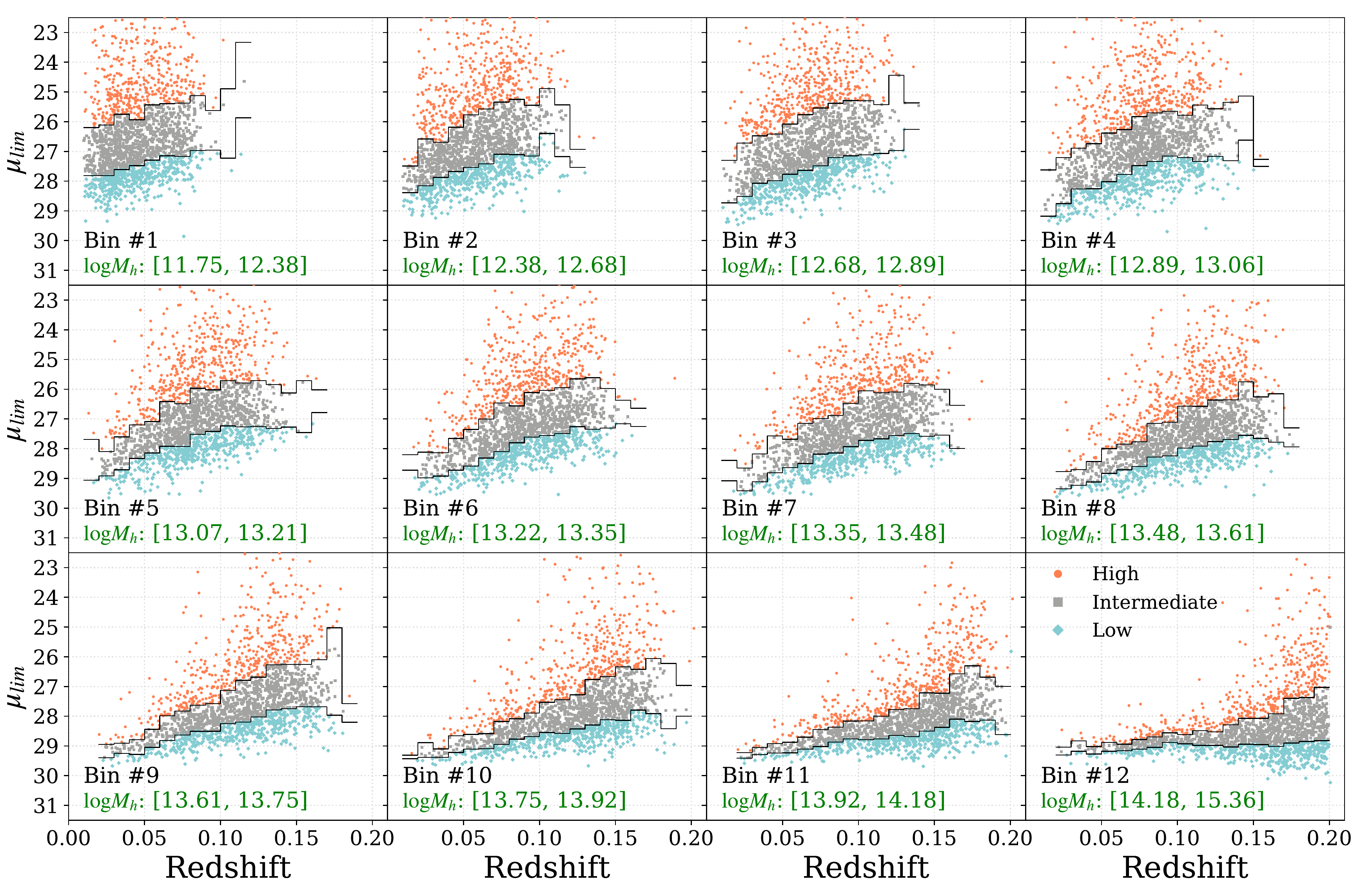}}
  \caption{The $\mu_{\rm lim}-$redshift diagrams for groups in different $M_h$ bins. The bin number and mass range of $M_h$ are given in the lower-left corner of each subplot. The solid step lines in each subplot represent the $25$th and $75$th percentiles of the $\mu_{\rm lim}$ in each redshift bin with a bin size of 0.01. The data points above the upper step lines are regarded as HC groups (red circle), those lying in the middle are regarded as IC groups (grey square), and those below the lower step lines are LC groups (blue diamond).}
  \label{fig:compactbin}
\end{figure*}

\subsection{Compactness of galaxy groups} \label{sec:compactness}
In this work, we use the group surface brightness as a proxy for group compactness. Following \citet{Hickson1982}, the surface brightness of a group system is defined as: 
\begin{equation}
\mu_{\rm lim} = -2.5\log{\left(\frac{\sum\limits_{i=1}^{N} 10^{-0.4 m_i}}{\pi \theta_{\rm G}^2}\right)}, \label{eq:mulim}
\end{equation}
\noindent where $\mu_{\rm lim}$ is the surface brightness averaged over the smallest circle enclosing $N$ galaxy members brighter than the magnitude limit within an angular radius of $\theta_{\rm G}$ in units of arcseconds, and $m_i$ is the apparent magnitude of the $i$th brightest member. 

To fairly define the compactness level of galaxy groups, we divide the groups into 12 $M_h$ bins with equal sample size and then plot $\mu_{\rm lim}$ against redshift for each $M_h$ bin in figure~\ref{fig:compactbin}. For each $M_h$ bin, we further divide groups into 20 redshift bins with a bin width of 0.01, and then calculate the $25^{\rm th}$ and $75^{\rm th}$ percentile values of $\mu_{\rm lim}$ distribution in each redshift bin. The solid lines in each panel of figure~\ref{fig:compactbin} show the separation lines of the $25^{\rm th}$ and $75^{\rm th}$ percentiles, respectively.  We see that there is a systematical redshift dependence of $\mu_{\rm lim}$ quartiles, where the higher redshift groups are biased toward higher $\mu_{\rm lim}$ values. This redshift bias is likely caused by the segregation effect of galaxies with different masses (see section~\ref{sec:discussion} for a detailed discussion). 

\begin{figure}
  \subfigure{
    \includegraphics[width=1.\columnwidth]{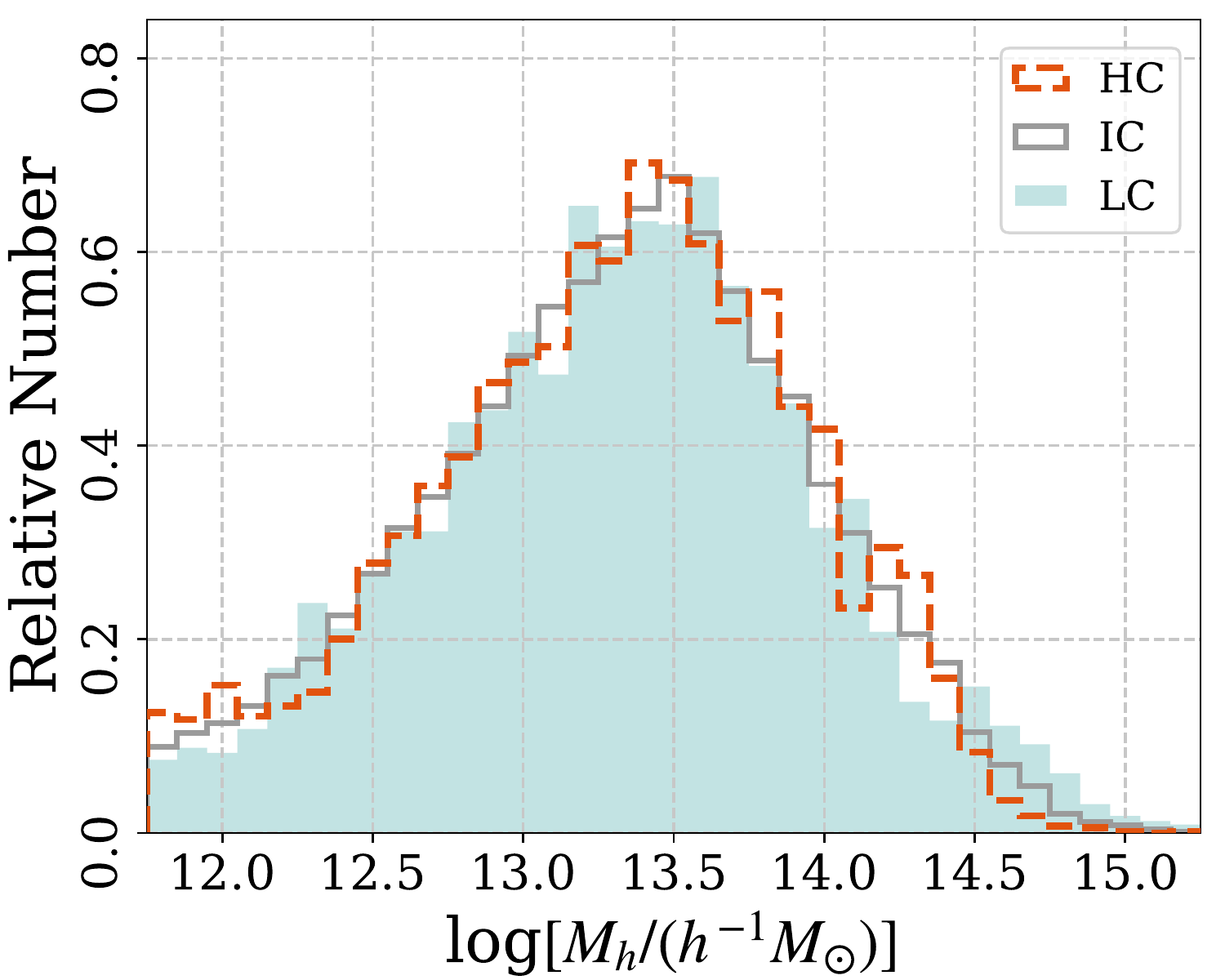}}
  \caption{The $M_h$ distributions for the HC (red dashed), IC (grey solid), and LC (blue filled) groups with a bin width of 0.1 dex.}
  \label{fig:MhBin}
\end{figure}

Here, we are not interested in well-defining the compactness of groups of galaxies with a quantitative $\mu_{\rm lim}$ criterion. On the contrary, our aim is to make a fair comparison of galaxy groups with different compactness while their other main physical features are well constrained. To do that, we consider the groups of each $M_h$ bin that lie in the top quartile to be high compactness (HC) groups, while those lying in the middle two quartiles are considered as intermediate compactness (IC) groups and those in the bottom quartile are considered to be low compactness (LC) groups. With this configuration, our HC and LC groups have the same $M_h$ and redshift distribution, and thus any differences in between these two samples would mainly correlate with the compactness of galaxy groups. In figure~\ref{fig:MhBin}, we show the $M_h$ distribution of the Y07 groups with different levels of compactness. As expected, these three different samples have nearly the same $M_h$ distributions, which peaks around $\sim 10^{13.5}h^{-1}M_{\odot}$.

Moreover, for the $M_h \lesssim 10^{13.5}h^{-1}M_{\odot}$ groups, the separation line of the top quartile is close to the criterion $\mu_{\rm lim} \sim 26.0$ mag arcsec$^{-2}$ that is used for defining the compact groups of galaxies \citep{Hickson1982}. That is to say, there might be a good correspondence between the classical compact groups and our low-mass HC groups. On the other hand, for high-mass systems ($M_h \gtrsim 10^{13.5}h^{-1}M_{\odot}$), the majority of the groups in the high compactness quartiles are actually not `compact' enough ($\mu_{\rm lim} > 26.0$ mag arcsec$^{-2}$).

\subsection{X-ray emission of galaxy groups} \label{sec:otx}
We perform X-ray detections for Y07 groups in the \textit{ROSAT} All Sky Survey (RASS) image data, which mapped the sky in the soft X-ray band ($0.1 - 2.4$ keV) with exposure time varying between 400 and 40000 seconds, depending on the direction in the sky. The resolution of the RASS images is 45 arcsec.

The algorithm we use to estimate the X-ray luminosity is similar to those of \citet{Shen..2008} and \citet{WangLei..2014}. As the first step, we determine the X-ray center of each group. Optically, the position of the central galaxy is usually taken to define the group center. Indeed, the X-ray peak position might be offset from the central galaxy \citep[e.g.,][]{DaiXY..2007, Koester..2007, Shen..2008, WangLei..2014}. We locate the RASS fields where the group reside, then apply a maximum-likelihood (ML) search algorithm to generate the X-ray source catalog with detection likelihood $L > 7$ in the $0.5-2.0$ keV band \citep{Voges..1999}. We match all the detected sources within a tolerance of $0.1 R_{180}$ from the central galaxies and regard the closest one as the X-ray center. The other X-ray sources within $0.1 R_{180}$ are regarded as part of the extended X-ray emission from this group, and those that lie in $R > 0.1 R_{180}$ region are regarded as a contaminated source that are not associated with the emission of intra-group medium. If none of the X-ray sources is detected within $0.1 R_{180}$, the central galaxy then is regarded as the X-ray center, most of our samples ($\sim 90 \%$) belong to such cases.

To obtain an unbiased estimate of the typical X-ray luminosity of a subset of galaxy groups, we stack each subsample with similar $M_h$ and group compactness. The stacking method is described in \citet{Shen..2008}. Briefly, we center each group image on the X-ray center. All the images are scaled to the same size in units of $R_{180}$ and the contaminated sources located at $R > 0.1 R_{180}$ are masked out. The X-ray background is estimated as the average flux inside an annulus with inner radius of $R_{180}$ and outer radius of $1.3 R_{180}$. By subtracting the background, a one-dimensional count rate profile as a function of radius is then obtained. 

For a stack of $N$ groups, the source count of photons, $N_{\rm S}$, that could be detected in RASS is given as
\begin{equation}
N_{\rm S} = \sum^{N}_{i=1} L_{\text{X},i} g\left(N_{\text{H},i}, T_{i}, z_{i}\right) t_{i},
\end{equation}
where $L_{\rm{X},i}$ is X-ray luminosity of the i$th$ group, $t_i$ is average RASS effective exposure time of  i$th$ group, and $g\left(N_{\rm{H},i}, T_i, z_i\right)$ is a coefficient converting bolometric X-ray luminosity to observed count rate as functions of redshift $z_i$, average galactic hydrogen column density $N_{\rm{H},i}$ in the line of sight and gas temperature within the i$th$ group. We assume that the gas metal abundance is assumed to be $0.3$ of the solar value, whereas $N_{\rm{H},i}$ is taken from the map of \citet{Dickey.Lockman1990} and the X-ray emission has a thermal spectrum with temperature $T$ obtained from the empirical $T-\sigma$ scaling relation of \citet{White..1997}. However, the slope of the $T-\sigma$ relation for small groups is  not necessarily the same as massive clusters,  and the gas temperature of low $M_h$ groups is even not well-constrained \citep[][and references therein]{Lovisari..2021}. Moreover, the conversion factor for hot gas with $T \lesssim 0.3$ keV (corresponding to $M_h \lesssim 10^{13} h^{-1} M_{\odot}$) becomes very  different for different X-ray spectral models (e.g., power law, thermal bremsstrahlung, Raymond-Smith plasma, and blackbody spectrum) over this range \citep{ROSAT}.   However, until there is more negative evidence,  there is reason to believe that the $N_{\rm{H}}$ and $T$ distributions for groups at fixed $M_h$ but with different compactness are similar, i.e., they have similar $g\left(N_{\rm{H},i}, T_i, z_i\right)$ distributions. Indeed, we have checked that the difference between the resulting X-ray luminosity of the LC groups and HC ones (we will show in figure~\ref{fig:LX}) is caused by the difference between their detected X-ray photon counts rather than  the specific values of the conversion factor $g\left(N_{\rm{H},i}, T_i, z_i\right)$. 

For simplicity, we use the conversion factors given by the table 2, 3, and 5 of \citep{Bohringer..2004}. For the groups with gas temperature below the lowest threshold of each table, we take use of the values in that row. Based on the assumption that the X-ray luminosities of the sources in a stack are similar, the weighted average X-ray luminosity, $L_{\rm X,S}$, of the stack can be given as
\begin{equation}
L_{\rm X,S} = \cfrac{N_{\rm S}}{\sum\limits^N_{i=1} g\left(N_{\text{H},i}, T_{i}, z_{i}\right) t_{i}}, \label{eq:stack}
\end{equation}

Following \citet{WangLei..2014}, the X-ray extension radius is set as a fixed value of $R_{\rm X} = 0.5 R_{180}$ and the source count rates are integrated into the X-ray extension radius $R_X$. By changing this fixed value from $R_{\rm X} = 0.3 R_{180}$ to $0.6R_{180}$, none of our main results is significantly impacted. Finally, to estimate the flux that is missed outside $R_{\rm X}$, we make a $\beta-$profile extension correction to make up the X-ray luminosity missed in the range $R_{\rm X} \le R \le R_{180}$ \citep[see][]{Bohringer..2000, Shen..2008, WangLei..2014} with $\beta = 2/3$ and core radius $R_c = 0.14 R_{180}$.  If $\beta$ is set as a free parameter in profile fitting, we find all the best fittings are in the range of $0.40-0.70$, but with large uncertainties. Therefore, we use a fixed value of $\beta$ to make the profile fitting and flux calibration.

\section{Sample: physical properties of group member galaxies}\label{sec:props}
\subsection{Quenched and Star-forming Galaxies} \label{sec:quench}
The $M_{\star}$ and star formation rate (SFR) of individual galaxies are directly retrieved from the MPA/JHU-DR7 release \citep{MPA-JHU,MPA-JHU-K} based on SDSS spectroscopy. There are $\sim 123,000$ out of ($140,930$) galaxies in our sample have been detected. These galaxies without match in MPA/JHU catalog are mainly caused by the fiber collision effect during the SDSS spectroscopy target.  In other words, their redshifts is obtained from other surveys rather than SDSS-DR7. 
Due to the need for determining central galaxies, the $M_{\star}$ of these unmeasured galaxies are obtained by the relation between stellar mass-light ratio and $(g-r)$ colors of \citet{Bell..2003}. Following \citet{LiCheng..2006}, the catastrophic $(g-r)$ color outside the $3\sigma$ ranges from the mean color-magnitude relations of both blue and red sequences is replaced by the mean color of the red or blue cloud, depending on their location on the diagram. 

Based on the bimodality of galaxies distributed in the $\text{SFR} - M_{\star}$ diagram, galaxies can be divided into star-forming and quenched population. We adopt the demarcation line suggested by \citet{Bluck..2016}, which is parallel to but 1 dex below the star formation main sequence:
\begin{equation}
\log{\text{SFR}}  = 0.73 \log{M_{\star}} - 1.46\log{h} - 8.3,
\end{equation}
where the reduced Hubble constant $h$ is included here to convert the units of $M_{\star}$ from $M_{\odot}$ used in \citet{Bluck..2016} to $h^{-2}M_{\odot}$ used here. We have also examined the results based on the demarcation line given by \citet{Woo..2013}, and find negligible changes.

\begin{figure}
  \centering
  \subfigure{
    \includegraphics[width=1.\columnwidth]{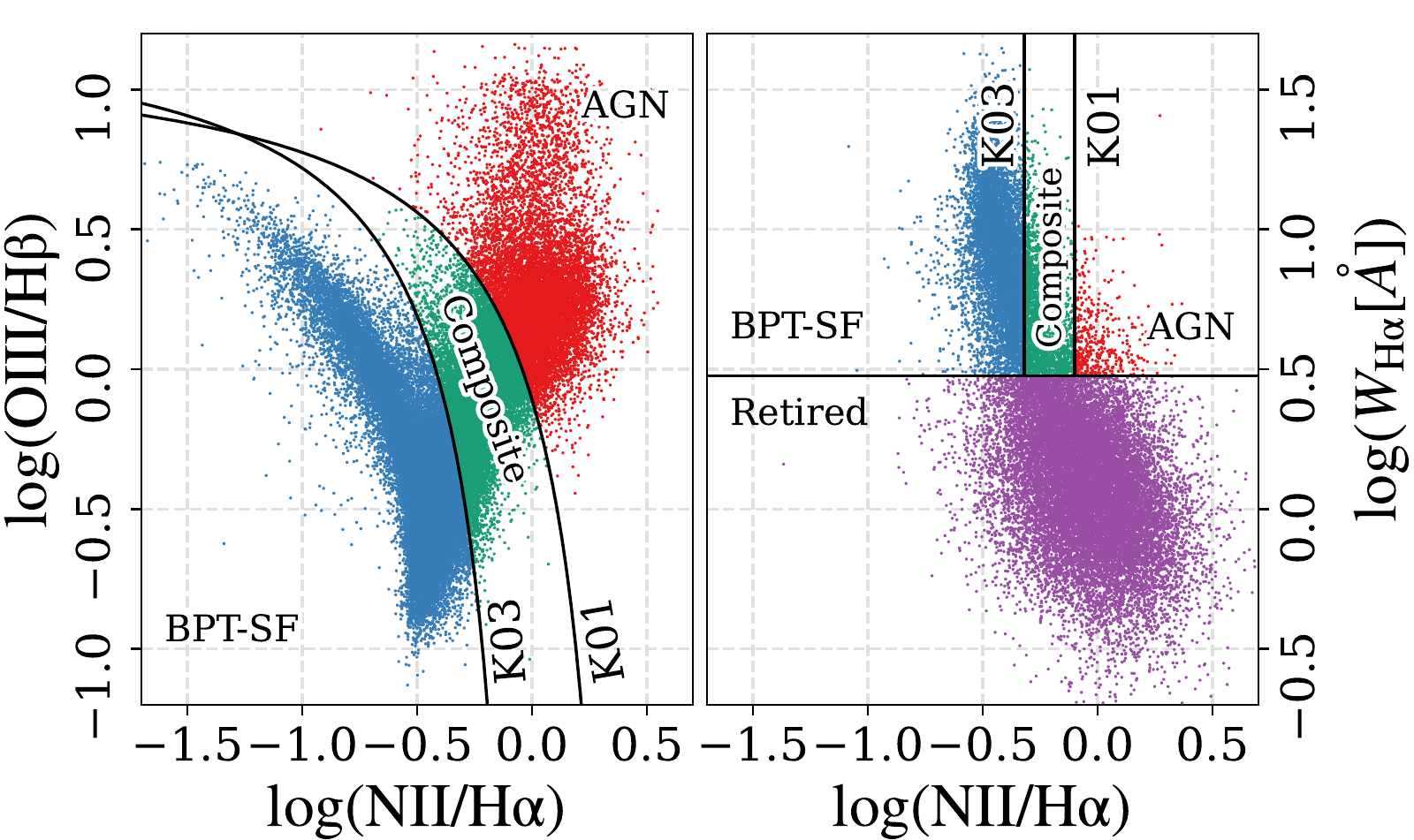}}
  \caption{Classification of spectral types for strong- and weak-line galaxies in Y07 groups. Left Panel: The BPT diagnostic diagram for strong-line galaxies with {\oiii}/{\Hb} versus {\nii}/{\Ha} emission-line ratios. The demarcation lines are taken from \citet[][K01]{K01} and \citet[][K03]{K03}. Right Panel: The WHAN diagram for weak-line galaxies in Y07 groups. The SF/composite/AGN demarcation lines are transposed from K01 and K03, respectively \citep{CF..2010}, and the retired/non-retired demarcation line is taken from \cite{CF..2011}.}
  \label{fig:agns}
\end{figure}

\subsection{AGN classification} \label{sec:agn}
\subsubsection{Optical AGNs} \label{sec:opagn}
The emission lines related to AGN classification are also taken from the MPA/JHU catalog.  The optical AGN are classified based on the BPT diagram \citep{BPT}, which is traditionally used to separate type II AGN from star-forming and composite galaxies. Here, we use the demarcation lines proposed by \citet{K01} and \citet{K03} (the left panel of figure~\ref{fig:agns}). Following \citet{Pasquali..2009}, the fluxes of the four emission lines are required to have a signal-to-noise ratio of $\rm S/N \ge 3$. This results in a sample of $\sim 50,000$ AGN out of $\sim 123,000$ strong-line galaxies. Besides, there are $\sim 28,000$ weak-line galaxies whose {\Ha} and {\nii} lines are both detected with an $\rm S/N \ge 3$, but either or both of {\Hb} and {\oiii} lines have a lower $\rm S/N$. Then, following \citet{CF..2010, CF..2011}, we use the WHAN method, which only requires {\nii}/{\Ha} and {\Ha} equivalent width ($W_{\Ha}$) to select weak-line AGN. We show the classification diagram for weak-line galaxies in the right panel of figure~\ref{fig:agns}. Note that the emission line fluxes have been corrected for intrinsic extinction based on the Balmer decrement and a dust attenuation curve of the form $\lambda^{-0.7}$ \citep{Charlot.Fall2000} by assuming an intrinsic ${\Ha}/{\Hb} = 2.86$ \citep{Osterbrock1989}.

\subsubsection{Radio-loud AGN} \label{sec:rdagn}
Radio-loud AGNs are obtained by matching our sample galaxies with the radio galaxy catalog of \citet{Best.Heckman2012}, which was established by cross-matching the MPA/JHU catalog with the NVSS \citep[National Radio Astronomy Observatory Very Large Array Sky Survey,][]{Condon..1998} and FIRST \citep[Faint Images of the Radio Sky at Twenty centimeters,][]{Becker..1995} surveys following the method of \citet{Best..2005}. The catalog is down to a flux density level of 5 mJy, which means that the catalog probes down to a radio luminosity of $L_{1.4 \rm GHz} = 10^{23} {\rm W} \cdot {\rm Hz}^{-1}$ at $z = 0.1$, we thus only consider the galaxies with $z \le 0.1$ \citep{WangEnci..2018} when investigating the radio-loud AGN fraction.

\begin{figure*}
  \centering
  \subfigure{
    \includegraphics[width=.86\hsize]{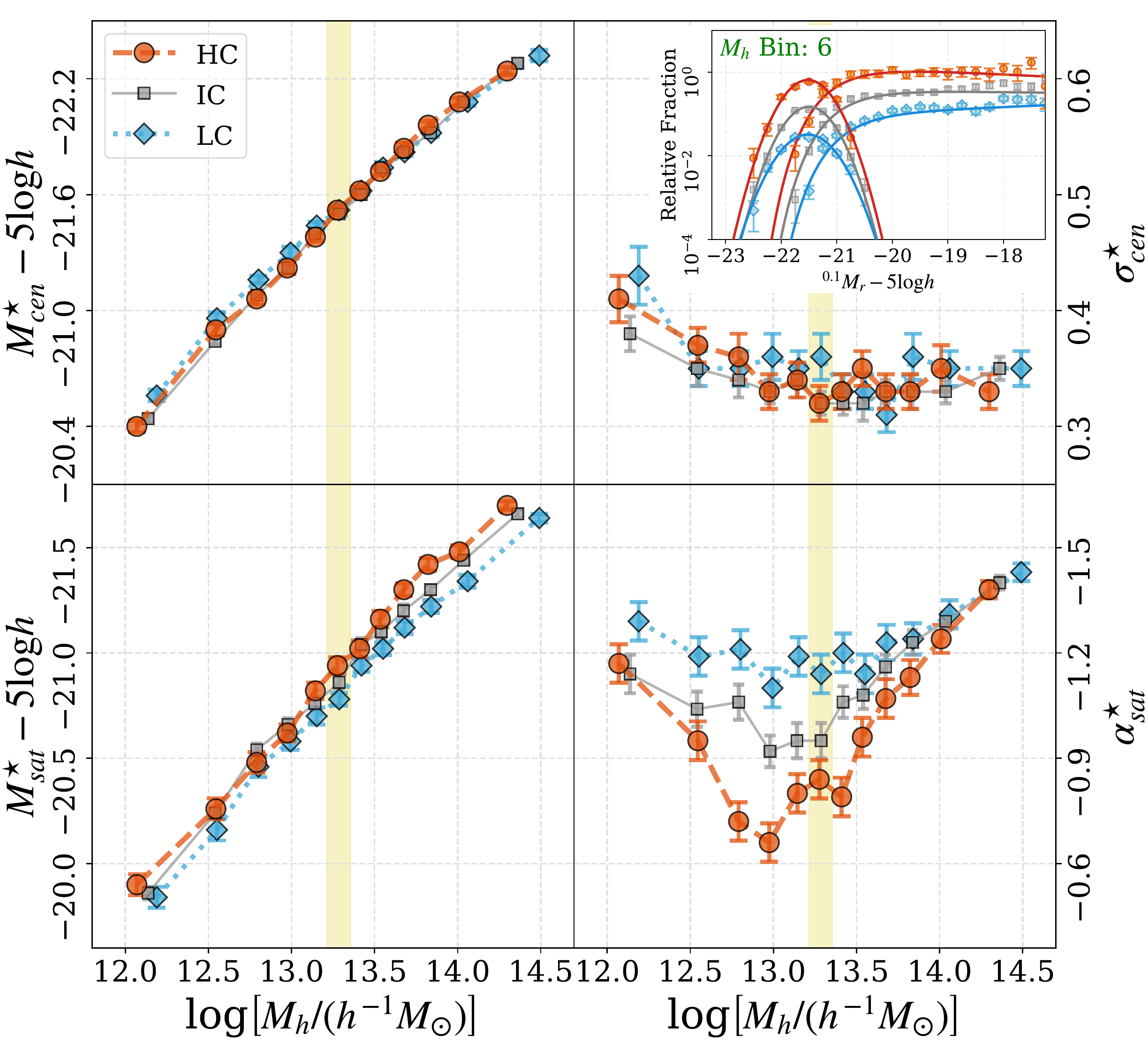}}
  \caption{The best-fit parameters ($M_{\rm cen}^{\star}$, $\sigma_{\rm cen}^{\star}$, $M_{\rm sat}^{\star}$, and $\alpha_{\rm sat}^{\star}$) of CLFs for the HC (red), IC (grey), and LC (blue) groups as a function of $M_h$. The vertical error bars show the errors of the parameter in each $M_h$ bin. The plot in the inset in the upper right corner demonstrates an example of a CLF fitting for the groups in $M_h$ Bin $= 6$ (shaded region) derived via both a nonparametric stepwise (symbols with error bars) and parametric ML estimator (solid lines), the fitting for each subsample is manually shifted for clarity.}
  \label{fig:clf3}
\end{figure*}

\section{The global properties of galaxy groups: dependence on group compactness} \label{sec:global}
In this section, we start with a detailed investigation of how the global properties of a galaxy group varied with the group compactness.

\subsection{CLFs} \label{sec:CLFs}
The dynamical state of a group system is directly related to the luminosity distribution of its galaxy members \citep{Martinez.Zandivarez2012}. As the first step in our investigation of how the group properties varied with group compactness, we use the CLF to describe the luminosity distribution of galaxies in groups with different levels of compactness at a given $M_h$. Following \citet{Yang..2008}, the total CLF can be modeled as the sum of the CLFs of central and satellite galaxies \footnote{In this section, we take the brightest galaxy in the group as the central galaxies instead of the most massive one. Tests have shown that both of the definitions yield indistinguishable results.}:
\begin{equation}
\Phi \left(^{0.1}M_{r}|M_h\right) = \Phi_{\rm cen} \left(^{0.1}M_{r}|M_h\right) + \Phi_{\rm sat} \left(^{0.1}M_{r}|M_h\right),
\end{equation}
The CLF of central galaxies is assumed to be a lognormal form:
\begin{equation}
\Phi_{\rm cen} \left(^{0.1}M_{r}|M_h\right) \propto \exp\left[-\frac{{\left(^{0.1}M_{r} - M_{\rm cen}^{\star}\right)}^2}{2 {\sigma_{\rm cen}^{\star}}^2}\right],
\end{equation}
where $M_{\rm cen}^{\star}$ is the expected value for the absolute magnitude of the central galaxy, and $\sigma_{\rm cen}^{\star}$ is the standard deviation of $M_{\rm cen}^{\star}$. The CLF of satellite galaxies takes a modified \citet{Schechter1976} function:
\begin{equation}
\begin{split}
\Phi_{\rm sat} \left(^{0.1}M_{r}|M_h\right) \propto &  10^{-0.4\left(\alpha_{\rm sat}^{\star}+1\right)\left(^{0.1}M_{r} - M_{\rm sat}^{\star}\right)} \\
 & \cdot \exp{\left[-10^{-0.8\left(^{0.1}M_{r} - M_{\rm sat}^{\star}\right)}\right]},
\end{split}
\end{equation}
where $M_{\rm sat}^{\star}$ is the characteristic magnitude and $\alpha_{\rm sat}^{\star}$ is the faint-end slope of the satellite galaxies. Here, we focus on the shape of these CLFs only, and the parameters ($M_{\rm cen}^{\star}$, $\sigma_{\rm cen}^{\star}$, $M_{\rm sat}^{\star}$, and $\alpha_{\rm sat}^{\star}$) of CLFs are obtained using ML estimator \citep{Tammann..1979}, where the binned CLFs are obtained via nonparametric stepwise maximum likelihood method \citep{Efstathiou..1988}. The results for each parameter as a function of $M_h$ are shown in figure~\ref{fig:clf3}, where we also plot an example of the fitting for the subsamples in $M_h$ Bin $= 5$ derived via both methods to show the consistency of the results. 

For central galaxies, we see that there is a strong dependence of CLF shape on $M_h$ values. However, for a given $M_h$, there is only a weak dependence of CLF parameters on group compactness. For high $M_h$ bins ($M_h \gtrsim 10^{13}h^{-1}M_{\odot}$), the three categories of galaxy groups with different compactness have almost identical $M_{\rm cen}^{\star}$ values. However, for low $M_h$ bins ($M_h \lesssim 10^{13}h^{-1}M_{\odot}$), LC groups tend to have slightly but systematically brighter $M_{\rm cen}^{\star}$ than the others. The scatter of central galaxies $\sigma_{\rm cen}^{\star}$ are consistent with each other inside uncertainties for all $M_h$ bins.

For the CLFs of satellite galaxies, we see not only a strong dependence on $M_h$, but also an obvious dependence on group compactness. The $M_{\rm sat}^{\star}$ of the LC groups is systematically fainter, while their faint-end slope, $\alpha_{\rm sat}^{\star}$, is more negative and hence contains more faint galaxy members than the HC groups. These two results are not independent of each other because there is a degeneracy between $M_{\rm sat}^{\star}$ and $\alpha_{\rm sat}^{\star}$ in modified Schechter function fitting when the total luminosity is fixed. Besides, at a fixed total luminosity, groups with more galaxy members (so that more negative $\alpha_{\rm sat}^{\star}$) might also be biased toward having higher $\mu_{\rm lim}$ values (because of the larger radius of the smallest enclosed circle). To check to what extent the difference of richness between the LC and HC groups at a given $M_h$ can explain the observed difference of $\mu_{\rm lim}$ values, we perform a simple test. More specifically, for a group with $N$ members at a given $M_h$, we take the median $\theta_G$ of the observed samples as satisfying the conditions as expected at the radius $R_{\rm m} (N, M_h)$. By assuming the total luminosity of each group at a fixed $M_h$ are the same, we thus can estimate the relative difference in $\mu_{\rm lim}$ distributions for the mock groups with the richness distribution assigned from HC and LC groups, respectively. Indeed, the disparity of the richness between the LC and HC groups at a given $M_h$ can only explain at most $\sim 20 \%$ of their $\mu_{\rm lim}$ difference.

\begin{figure}
  \centering
  \subfigure{
    \includegraphics[width=1.\columnwidth]{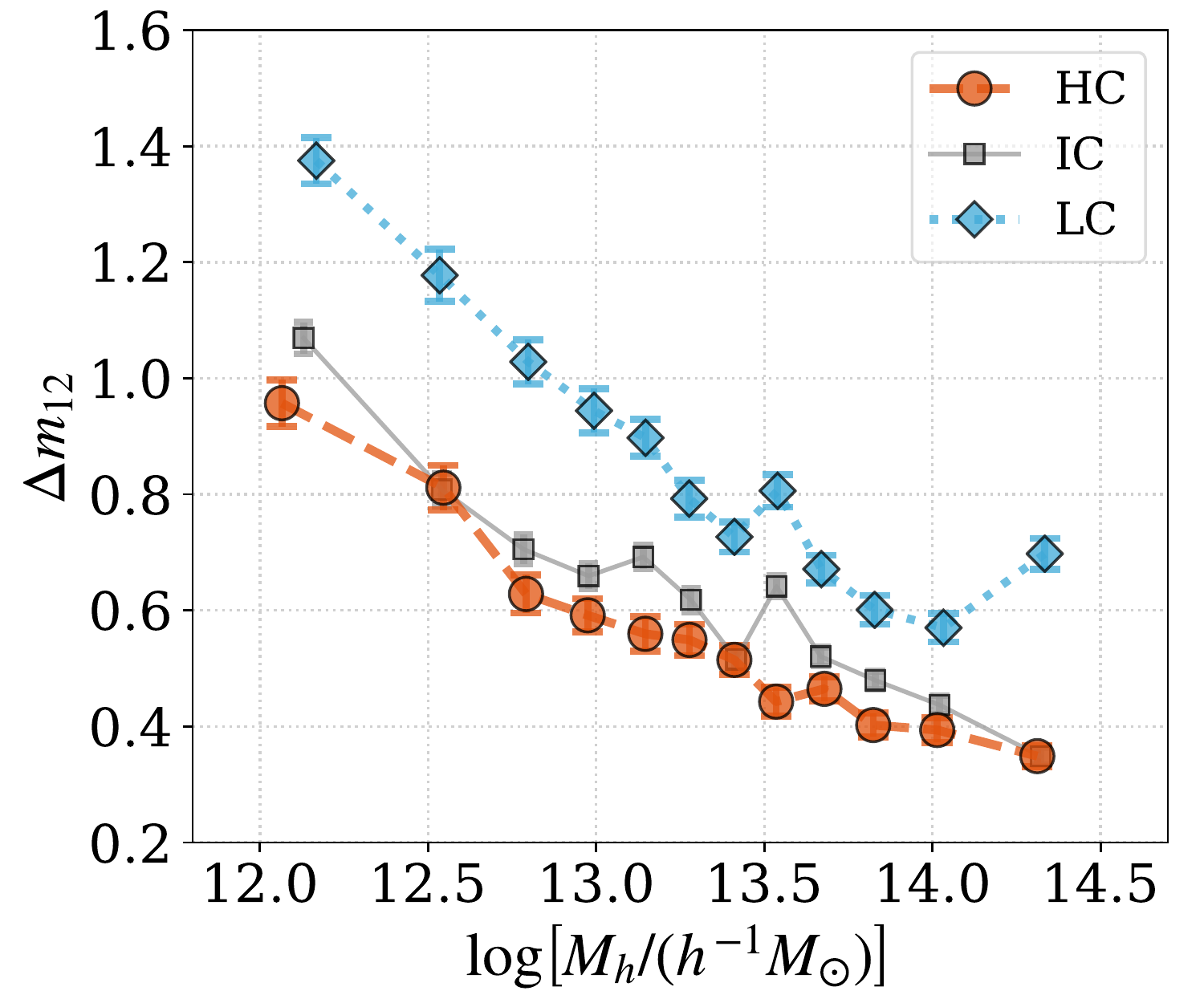}}
  \caption{The median $\Delta m_{12}$ as a function of $M_h$ for the HC (red), IC (grey), and LC (blue) groups. The vertical error bars show the errors of the median $\Delta m_{12}$ in each $M_h$ bin. Note that the HC, IC, and LC groups in each $M_h$ bin have been controlled to have similar redshift and richness distributions.}
  \label{fig:m12}
\end{figure}

\subsection{Luminosity gap: $\Delta m_{12}$} \label{sec:m12}
The CLF results presented in the last section imply that the HC groups (especially for those with $M_h< 10^{13.5}h^{-1}M_{\odot}$) are biased to have fainter central galaxies and brighter satellites. Therefore, the natural conclusion is that the HC groups would be biased to have lower magnitude gap  $\Delta m_{12}$. In this subsection, we investigate in detail how $\Delta m_{12}$ is correlated with the group compactness.

Based on the principle of order statistics, \citet{Shen..2014} have shown that $\Delta m_{12}$ is mainly determined by group richness, where groups with more members would be naturally biased to have smaller $\Delta m_{12}$. On the other hand, for groups with a given $M_h$, $\Delta m_{12}$ has long been conjectured to reflect their dynamical states \citep[e.g.,][]{More..2012, Paranjape.Sheth2012, Tal..2012}. This argument is based on the fact that the dynamic friction drives massive major pairs to merge more quickly than other mergers within the same halo. The massive satellites would be coalesced by the central galaxies earlier, which drives $\Delta m_{12}$ to increase with time. However, this simple scenario might not be enough to account for the observed $\Delta m_{1,2}$ of galaxy groups since all galaxy systems are in hierarchical merging and a recent merging event of group systems may significantly change $\Delta m_{12}$ \citep{Trevisan.Mamon2017}.

To perform a fair comparison of $\Delta m_{12}$ between the groups with different compactness, we do not compare their $\Delta m_{12}$ distribution in a straightforward way. We match the LC (IC) groups to the HC groups at the same $M_h$ bin against the confounding covariates, $X = \{ N, z \}$. We search for the counterparts within a distance of 1 scaled by a tolerance of $\{ \Delta N, \Delta z \} = \{ 0, 0.01 \}$ in the multi-dimensional space formed by $X$. We select the closest one (two, for IC groups) of them not selected yet. If all of the satisfied targets have been selected before, we pick the closest one again. If none of the targets satisfy the criterion, this HC group will be discarded. 

Figure~\ref{fig:m12} shows the median $\Delta m_{12}$ for the HC, IC, and LC groups in each $M_h$ bin. Here we see that $\Delta m_{12}$ decreases with increasing $M_h$, mostly due to the richness being generally higher for massive groups. On the other hand, the LC groups exhibit systematically larger $\Delta m_{12}$ than the HC groups. This result confirms the conclusion we have obtained from the CLFs in the last section, which is also consistent with the fact that classical compact groups are generally comprised of several relatively more luminous/massive galaxies close in proximity to one another. 

This result also raises a question on the dynamical status of the HC groups. Previous studies have suggested that galaxy groups show similar dynamical features (velocity dispersion, mean harmonic separation, and dynamical mass) in the late collapse and dynamical friction phases, although the time interval between these two phases is up to several gigayears \citep{Mamon1993, Mamon2007, Paper2}. Hence, there are two possibilities for these observed HC groups, one is that the HC groups are dynamical young systems in the late forming stage, and the other is that they are dynamical old systems in late merging stage. Obviously, the systematically smaller $\Delta m_{12}$ for the HC groups is much more likely to favor the former scenario. To further verify this hypothesis, we investigate the X-ray properties of groups with different compactness in the next section.

\begin{figure}
  \centering
  \subfigure{
    \includegraphics[width=1.\columnwidth]{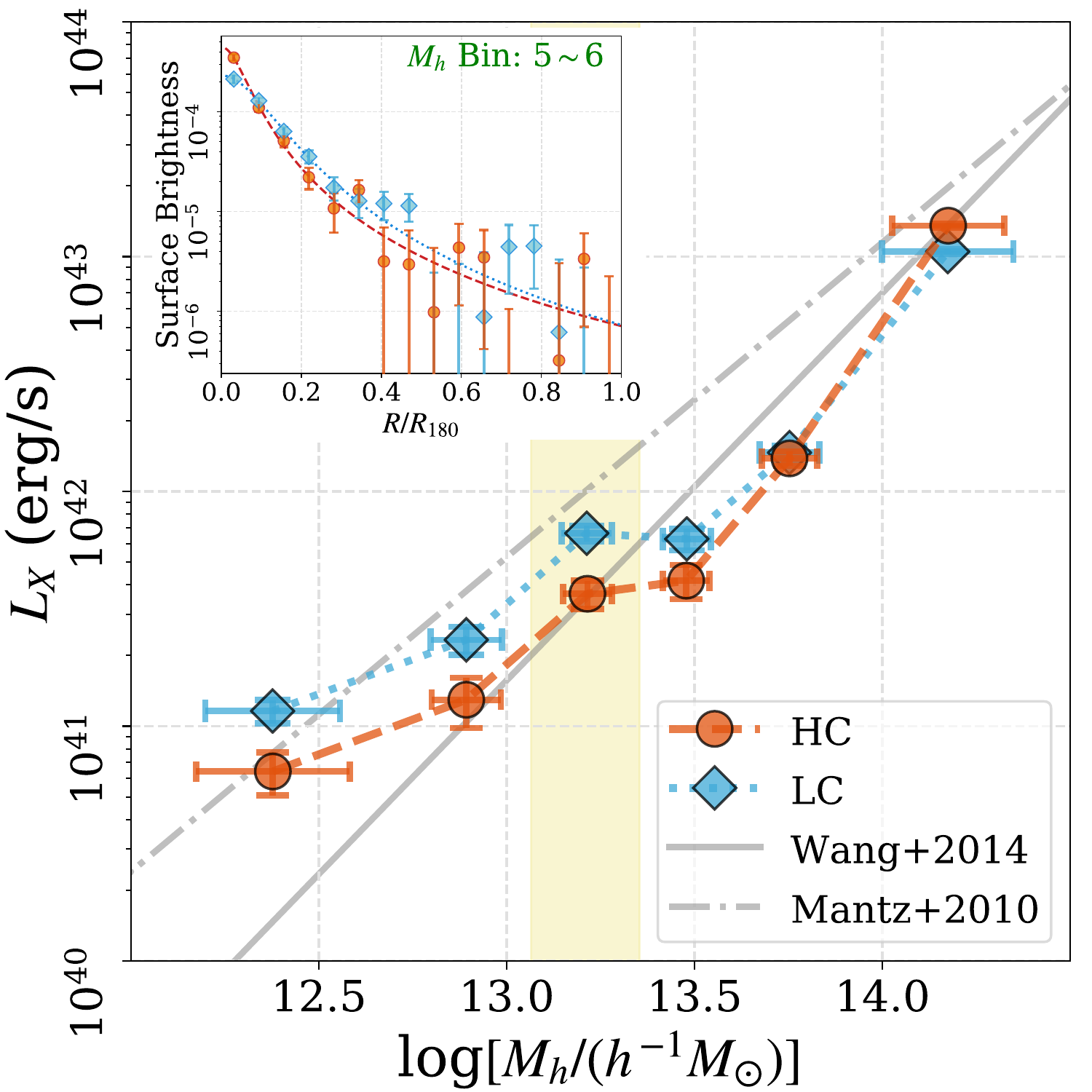}}
  \caption{The stacked X-ray luminosity as a function of $M_h$ for the HC (red) and LC (blue) groups. The vertical error bars show the errors transported from the uncertainties of the count rate and the horizontal error bars indicate the median deviation of the $\log{\left[M_h/(h^{-1}M_{\odot})\right]}$ distribution in each bin. The solid and dashed-dotted lines represent the results obtained by \citet{WangLei..2014} and \citet{Mantz..2010}, respectively. The plot in the inset in the upper left corner presents an example of surface brightness profiles (symbols with error bars) with $\beta-$model fittings (lines) for stacks of HC (red) and LC (blue) groups in $M_h$ Bin $= 5 - 6$, corresponding to the shaded region in the main plot.}
  \label{fig:LX}
\end{figure}

\subsection{X-ray Luminosities of Galaxy Groups} \label{sec:x-ray}
In this section, we investigate how the average X-ray luminosity of galaxy groups depend on the group compactness. To increase the statistical significance of our results, we rebin each pair of the adjacent $M_h$ bins in the following. We stack the groups of similar $M_h$ and compactness independently. In figure~\ref{fig:LX}, we present the average X-ray luminosity given by equation~\ref{eq:stack} as a function of $M_h$ for the stacks of each subsample and show an example surface brightness profiles for the stacks of the HC and LC groups in $M_h$ Bin $= 5 - 6$. As one can see, the $S/N$ of the surface brightness profile is sufficiently good and can be well represented by the $\beta-$model.

Broadly, the X-ray luminosity of galaxy groups has a good correlation with $M_h$, in agreement with the results of \citet{Mantz..2010} and \citet{WangLei..2014}. Note that the X-ray luminosity and halo mass are defined in a different way in \citet{Mantz..2010}, we transform both of the quantities using the method following \citet{WangLei..2014}. For the groups with $M_h \lesssim 10^{13.5}h^{-1}M_{\odot}$, the stacked X-ray luminosity of the LC groups is significantly brighter than the HC groups at a given $M_h$. 

X-ray emission of clusters/groups of galaxies has long been considered to reflect the dynamical status of group systems. X-ray selected groups generally have lower late-type fractions than typical optically-selected ones and are known to be biased toward relaxed systems, particularly in the low-mass regime \citep[e.g.,][]{Osmond.Ponman2004, Eckert..2011, Wen.Han2013}. Therefore, the systematically lower $L_{\rm X}$ for the HC groups, again, implies that they are potentially in a dynamical young status. To find further evidence of the correlation between the compactness of groups and their dynamical status, we investigate the star formation properties of individual galaxies living in groups with different compactness in the following section.

Besides the dynamical status, the X-ray emission of the intragroup medium is also known to be correlated with the feedback process of the radio AGN of the central galaxy \citep[e.g.,][]{Shen..2008}. We will present a detailed study of the AGN phenomena in groups of galaxies with different compactness in section~\ref{sec:agns}.

\begin{figure*}
  \centering
  \includegraphics[width=.49\hsize]{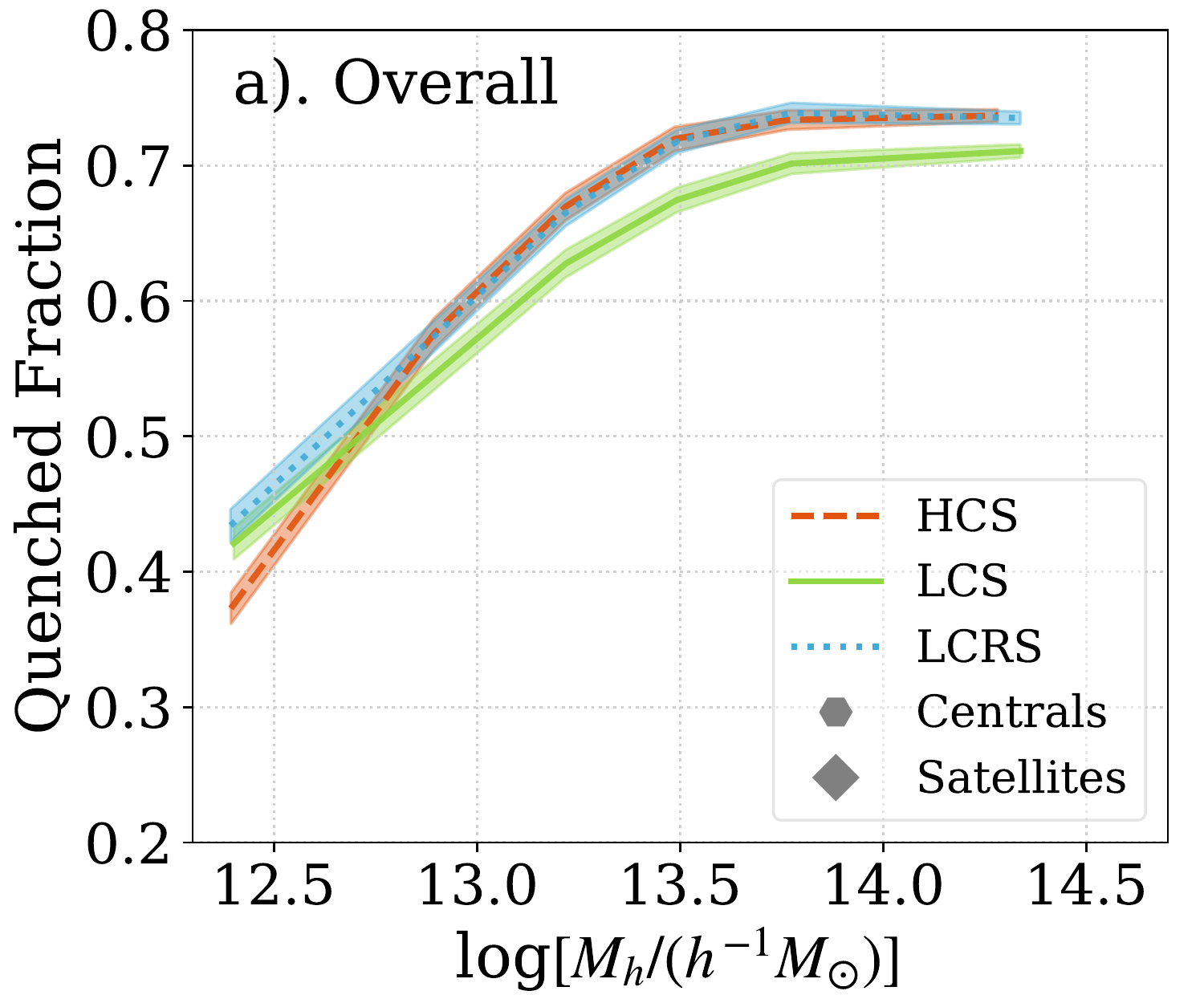}
  \includegraphics[width=.49\hsize]{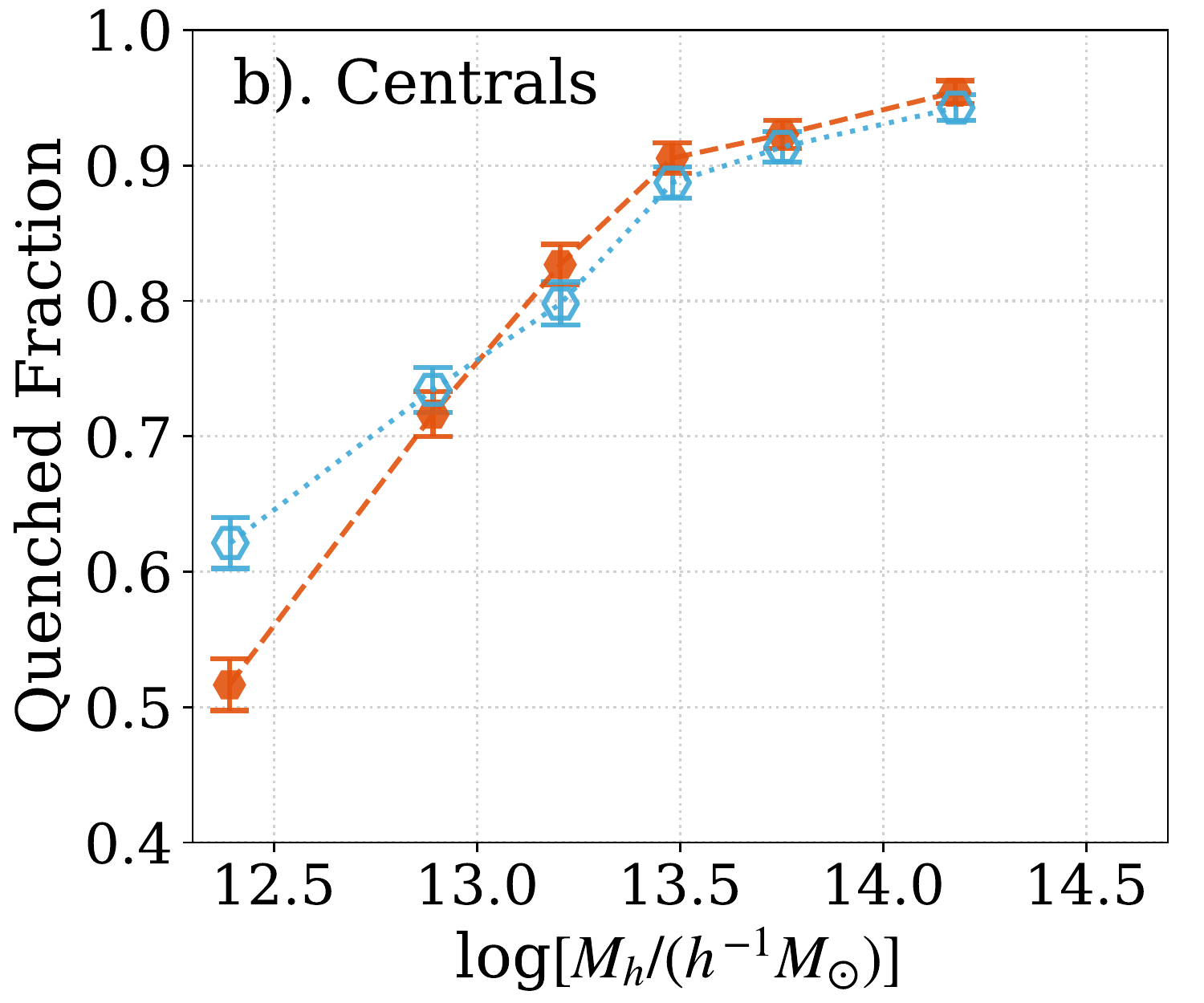}
  \includegraphics[width=.49\hsize]{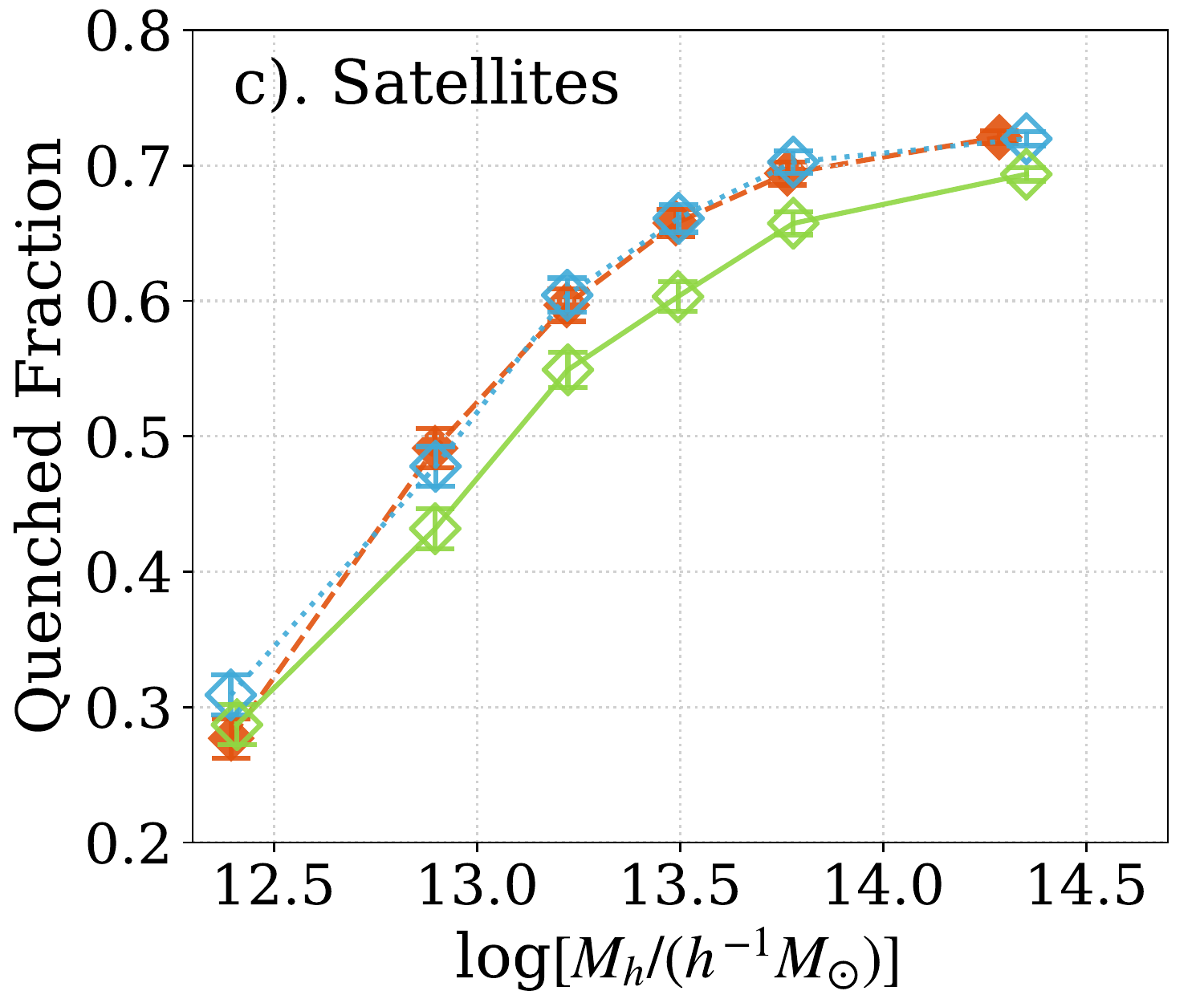}
 \caption{The quenched galaxy fraction for (a). overall, (b). central, and (c). satellite galaxies of the HCS (red), LCS (green), and LCRS (blue). Note that the central galaxies of LCS and LCRS are exactly the same; thus, we do not plot the results for the LCS in panel (b). In each panel, the errors in each bin represent the 68 percent confidence limits estimated using 1000 bootstrap resamplings.}
  \label{fig:fq}
\end{figure*}

\section{The properties of group member galaxies: dependence on group compactness} \label{sec:properties}
\subsection{Stellar Population of Galaxies} \label{sec:quenched}
We start by investigating the quenched fraction, which is defined as the fraction of galaxies that are quenched according to the aforementioned criterion of the HC and LC groups. Since our aim is to perform a fair comparison between the galaxies within the HC and LC groups at a given $M_h$, the bias due to different CLFs for these compared samples should be mitigated. The control samples for the HC and LC group galaxies are derived following the same method as described in section~\ref{sec:m12}. Briefly, we match the central and satellite galaxies of the LC groups to those in the HC groups against the confounding covariates, $X_{\rm cen} (\text{or }X_{\rm sat}) = \{ \log{M_{\star}}, z \}$ with a tolerance of $\{ \Delta \log{M_{\star}}, \Delta z \} = \{ 0.2, 0.01 \}$, respectively. The control samples derived in this way are referred to as control samples in the HC groups (HCS) and LC groups (LCS), respectively.

In figure~\ref{fig:fq}, we first plot the quenched fraction as a function of $M_h$ for the overall, central, and satellite population in three different panels, where the green solid and red dashed-dotted lines represent LCS and HCS, respectively. As a global trend, the quenched fraction increases systematically with $M_h$ for each subsample.  Since SDSS MGS is a flux-limited sample, the higher $M_h$ systems are biased toward higher redshift than the lower $M_h$ ones. Therefore, galaxies within higher $M_h$ groups are on average more massive, so they have a higher quenched fraction (known as mass quenching). On the other hand, at a given $M_{\star}$, galaxies in higher $M_h$ groups are also known to have a higher quenched fraction (known as halo quenching).  Therefore, the systematically higher quenched fraction of galaxies in higher $M_h$ bins we show in figure~\ref{fig:fq} is a combined result of both mass and halo quenching.

Besides the least massive $M_h$ bin (we will discuss it later), we find that the relative quenched fraction for all galaxies in the LCS is systematically lower than that in the HCS, which mainly results from their systematically lower quenched fraction for satellite galaxies (the lower panel of figure~\ref{fig:fq}). This result is in agreement with earlier findings that the galaxies in classical compact groups are on average redder than their counterparts living in noncompact groups \citep[e.g.,][]{Coenda..2012, Coenda..2015}. On the other hand, recent studies \citep[e.g.,][]{vandenBosch..2008, Wetzel..2012, Kauffmann..2013, WangEnci..2018} have found a dependency of galaxy quenching on halo-centric distance, where galaxies in central regions tend to be more quenched than their counterparts in the outskirts. In our work, galaxies in the LCS and HSC have been controlled to have the same $M_h$ so that have the same $R_{180}$. Therefore, the satellite galaxies of the HCS, by definition, naturally have systematically smaller  $R/R_{180}$ distributions than those of the LCS. That is to say, the systematically higher fraction of quenched galaxies in the HCS that we have found might be a result of their smaller average halo-centric distance. 

To test this hypothesis, we construct another control sample of LC group galaxies, the central population of which remains unchanged, by matching the satellite population to those in the HCS with the confounding covariates, $X_{\rm sat} = \{ \log{M_{\star}}, z, R/R_{180} \}$, scaled by a tolerance of $\{ \Delta \log{M_{\star}}, \Delta z, \Delta R/R_{180} \} = \{ 0.2, 0.01, 0.025 \}$. The control sample of LC groups derived in this way is referred to as the R-controlled sample in LC groups (LCRS). We plot the quenched fraction of galaxies for the LCRS as the blue dotted lines in each panel of figure~\ref{fig:fq}. 

We see that, for galaxies in the LC groups, when $R/R_{180}$ is further controlled, the quenched fraction of satellite galaxies has systematically increases. The global quenched fraction of galaxies in the LCRS is now in excellent agreement with the galaxies in the HCS. This result implies that the halo quenching effect might be a faster process. In section~\ref{sec:global}, from the global properties of groups (e.g., magnitude gap, X-ray luminosity), we have argued that the HC groups are likely to be in the early stage of group merging (or halo formation). Here, we see that although the HC groups are more likely to be recently formed, their members already show significant halo-centric dependence, i.e., galaxies in central regions of the halo are more likely to be quenched than in the outskirts.

On the other hand, although galaxies in the HCS and LCRS have been controlled to have the same $R/R_{180}$ distributions, galaxies in the HCS actually still have systematically higher local density than those in the LCRS. The higher local density means that they are more likely to show a galaxy-galaxy interaction effect. Such an effect is manifested in the lowest $M_h$ bin, where the local galaxy density (or $\mu_{\rm lim}$) of the HCS peaks (see also figure~\ref{fig:compactbin}). For the least massive $M_h$ bin where the halo quenching mechanism might not be very significant, we see a local galaxy-galaxy interaction effect, the galaxies in the HCS have systematically lower quenching fraction (because of enhanced star formation) than the LCS. Besides this bin, we see that the central galaxies in the HCS show a marginal increase in their proportion of quenched galaxies, this result might be affected by the magnitude measurement bias remaining in the pipeline, which we will discuss this in more detail in section~\ref{sec:magbias}.

\begin{figure*}
  \centering
  \includegraphics[width=.49\hsize]{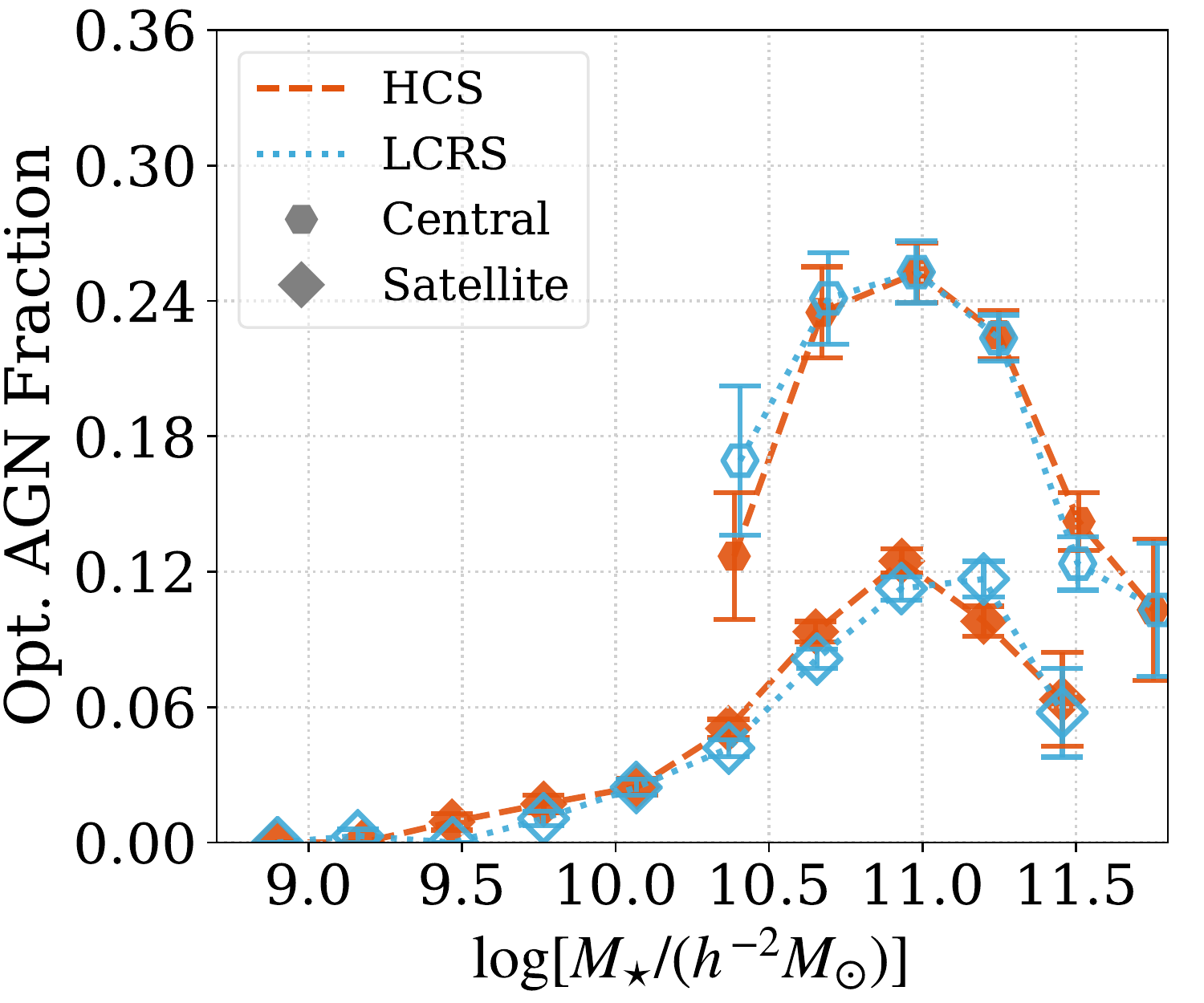}
  \includegraphics[width=.49\hsize]{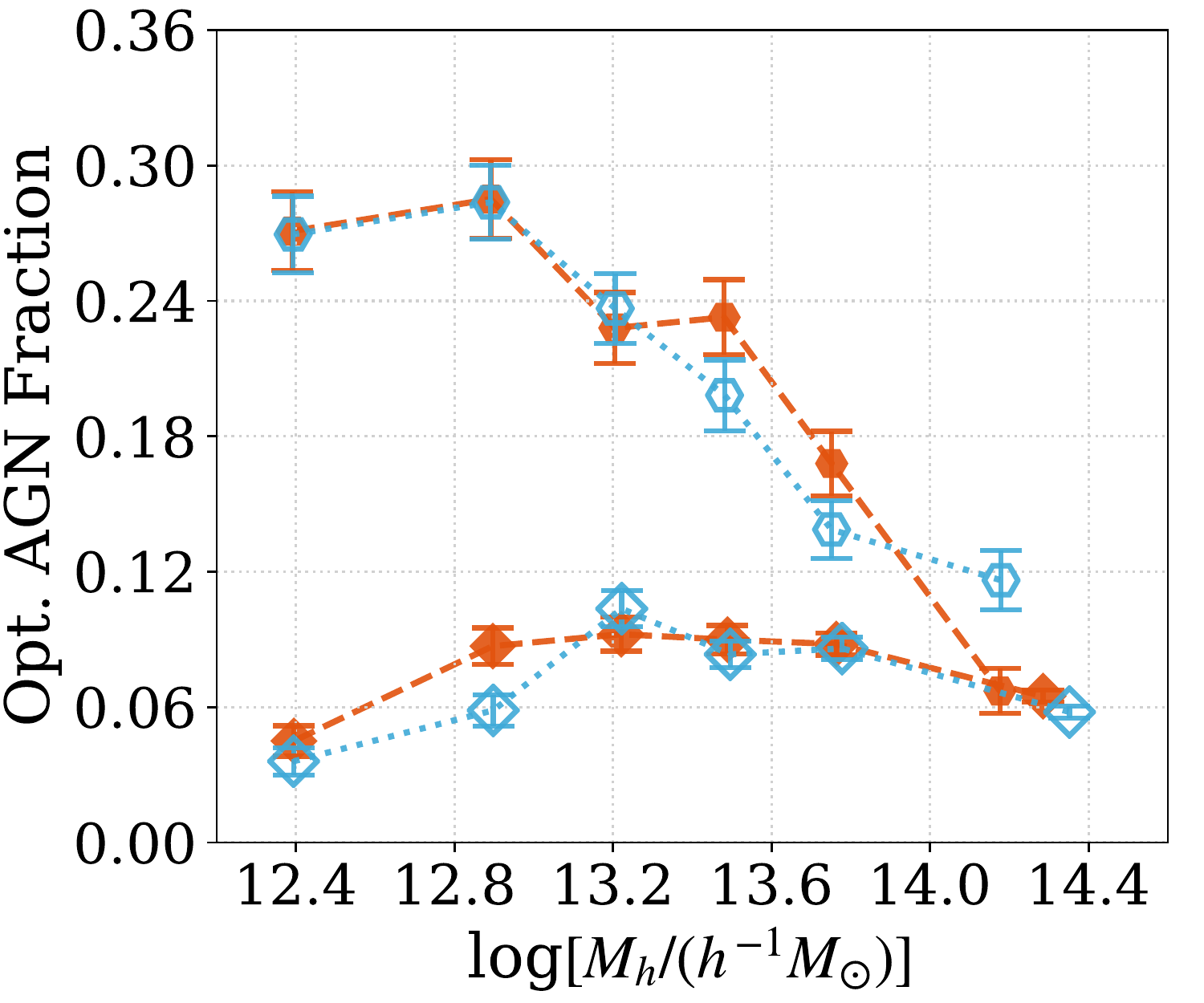}
  \includegraphics[width=.49\hsize]{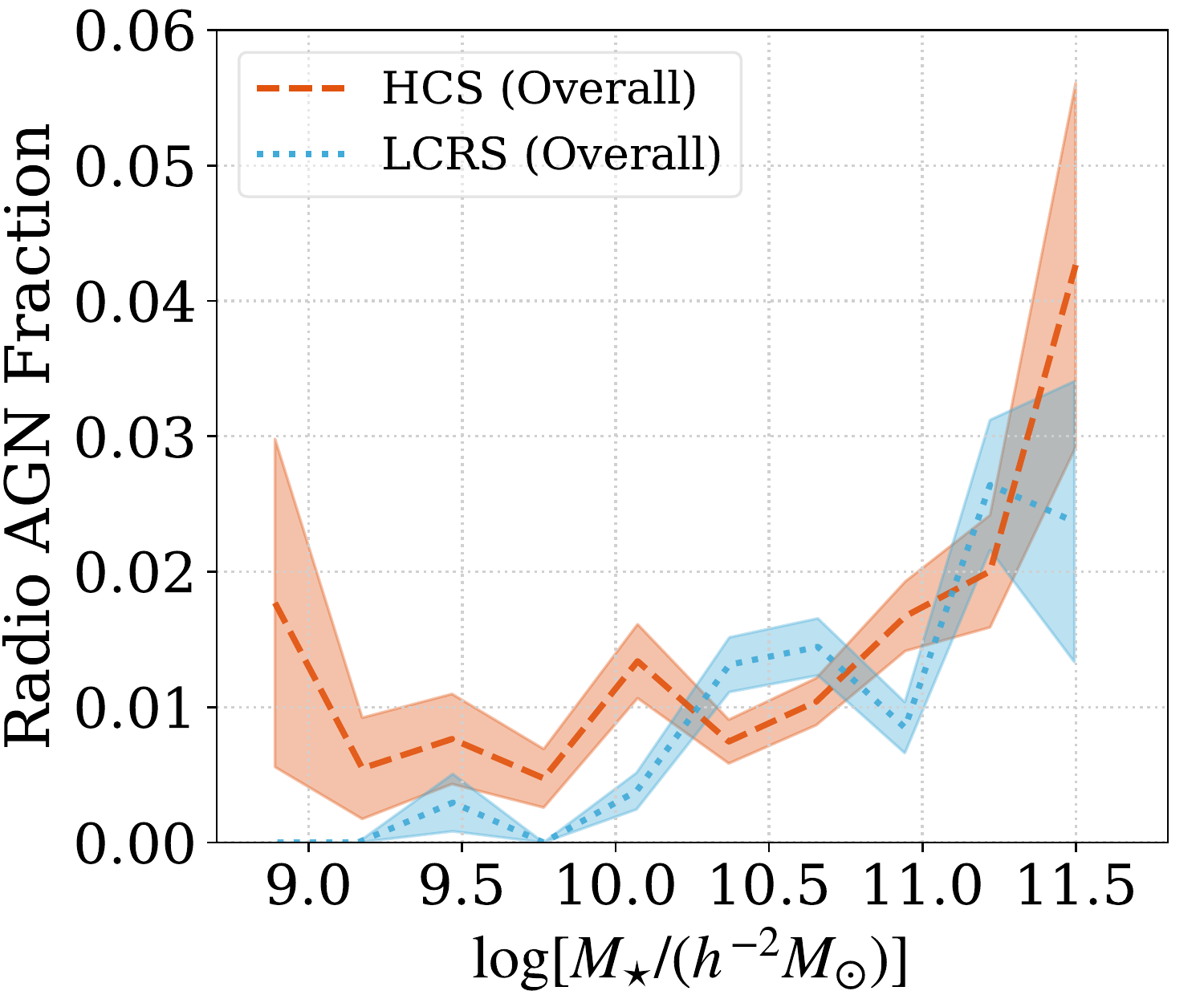}
  \includegraphics[width=.49\hsize]{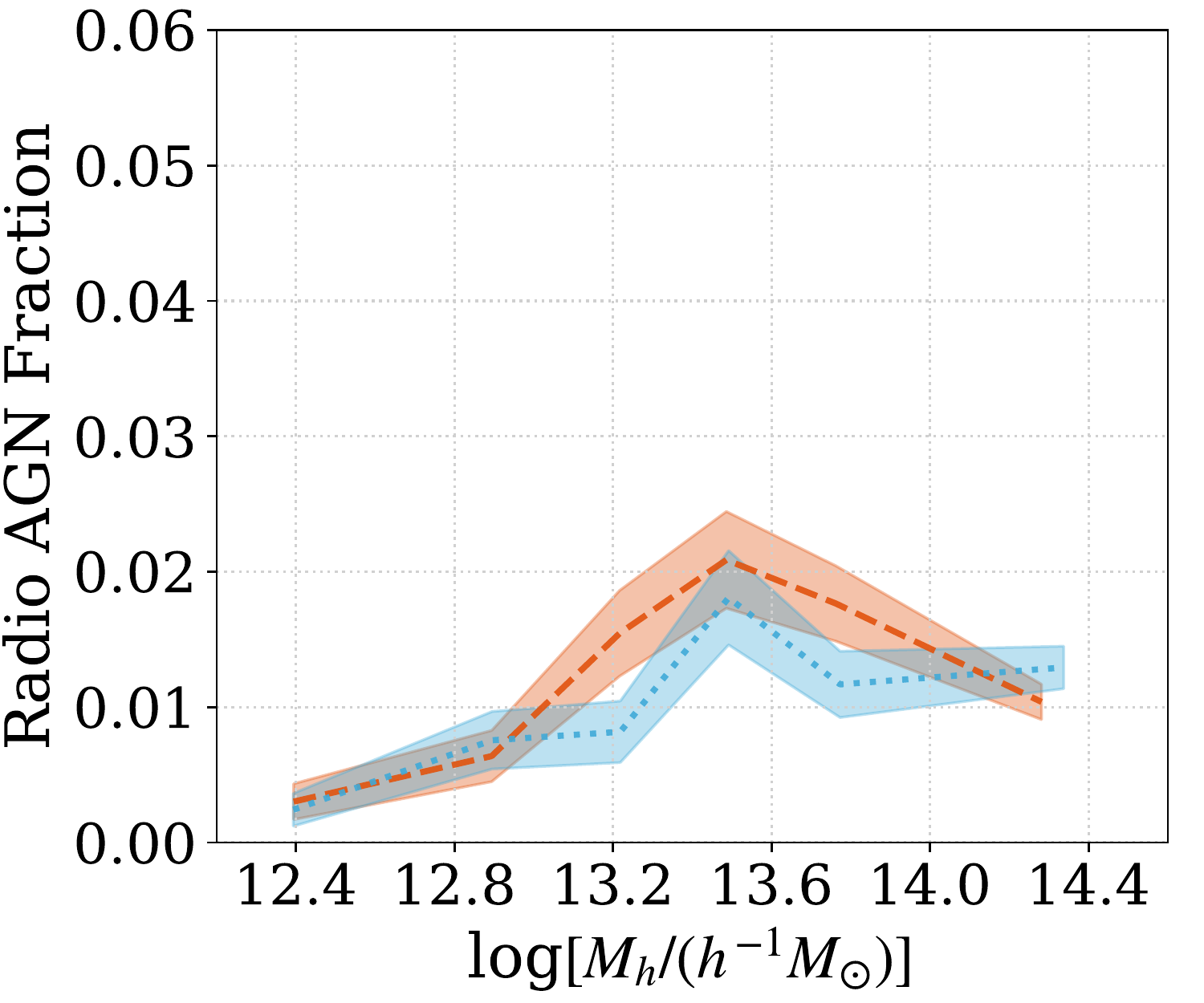}
  \caption{Upper Panel: The optical AGN fraction for central (hexagon) and satellite (diamond) galaxies of the HCS (red) and LCRS (blue), respectively. Lower Panel: The radio-loud AGN fraction for all galaxies of the HCS and LCRS. In each subplot, the errors in each bin represent the 68 percent confidence limits estimated using 1000 bootstrap resamplings.}
  \label{fig:fagn}
\end{figure*}

\subsection{AGN Properties} \label{sec:agns}
In the last section, we have shown that galaxies in the HCS have established a halo-centric relation as galaxies in the LCS. Here, we further explore whether the AGN phenomena of galaxies have been influenced by the formation phase of galaxy groups. To better control the variables, we focus on the HCS and LCRS only. 

We show the fraction of optical AGN in central and satellite populations for the HCS and LCRS in the upper two panels of figure~\ref{fig:fagn}, where the left panel shows the fraction as a function of halo mass $M_h$, while the right panel shows that as a function of their stellar mass $M_{\star}$. We see that the global trend of optical AGN phenomena shows a complicated dependence on $M_h$ and $M_{\star}$, which has been studied in detail by many previous works \citep[e.g.,][]{Pasquali..2009, WangEnci..2018, ZhangZW..2021}. We give a brief overview here. First, the optical AGN phenomena peak around $M_{\star} \sim 10^{10.5-11} h^{-2} M_{\odot}$, and decreases with increasing $M_h$ at $M_h \gtrsim 10^{12}h^{-1}M_{\odot}$. The central population has a higher fraction of optical AGNs than satellites at a given $M_{\star}$, which is due to the fact that the former live in the groups with lower $M_h$. On the other hand, the central galaxies show a higher average optical AGN fraction than satellites at a given $M_h$ for $M_h \lesssim 10^{14}h^{-1}M_{\odot}$ because the former contains more $M_{\star} \sim 10^{10.5-11}h^{-2}M_{\odot}$ galaxies. In addition to these known overall trends, for subsamples of the HCS and LCRS, we do not see any systematic differences between the HCS and LCRS. 

For radio-AGN phenomena, owing to the fact that radio-loud AGN are not very abundant, and \citet{WangEnci..2018} showed that there is no difference in radio properties between central and satellite galaxies at given $M_h$ and $M_{\star}$, we do not separate the galaxies into central and satellite populations here. In the lower panels of figure~\ref{fig:fagn}, we show the radio-loud AGN fraction for all galaxies in the HCS and LCRS as a function of $M_{\star}$ (left) and $M_h$ (right), respectively.  We also do not see any significant differences between the galaxies in HCS and LCRS. However, it is worth mentioning that the radio-loud AGN fraction of small satellites in the HCS sample seems systematically higher than that of their counterparts in the LCRS. Such a difference is interesting but suspicious since we know that the radio-loud AGN phenomena is linked with super-massive black holes which generally live in massive galaxies. This suspiciously higher radio-loud fraction at the low $M_{\star}$ end might come from the contamination during the identification of radio-loud AGN. An FR-II galaxy have two radio lobes separated out to a distance of $\sim 100$ kpc, which is very likely to contaminate its neighbors if this radio galaxy is located in a higher density region like the HC groups \citep{Best..2005}. A detailed study of the origin of this systematical difference is beyond the scope of this paper and deserves to be carried out in the future.

In summary, we find no systematic differences in AGN phenomena between the galaxies in the HCS and LCRS when their halo-centric effect has also been controlled. This result is consistent with an earlier finding that there is no dependence on optical nuclear activity within the local environment of galaxies \citep[e.g.,][]{Argudo..2016}.

\begin{figure}
  \centering
  \subfigure{
    \includegraphics[width=1.\columnwidth]{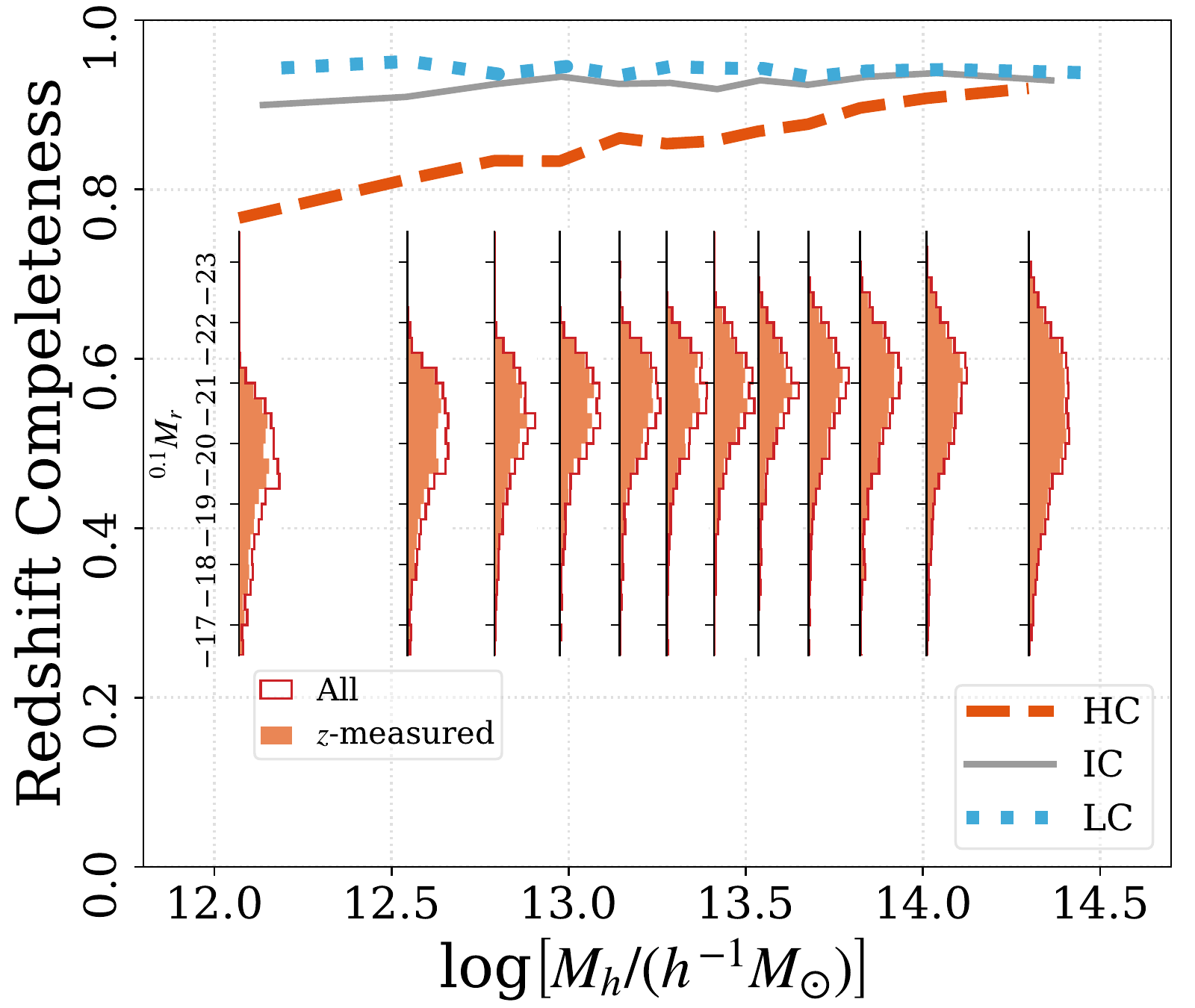}}
  \caption{Redshift completeness for galaxies living in the HC (red dashed), IC (grey solid), and LC (blue dotted) groups as a function of $M_h$. The insets represent the $^{0.1}M_{r}$ distributions for the HC group galaxies with (filled) or without (open) redshift measurement in each $M_h$ bin.}
  \label{fig:zcomp}
\end{figure}

\section{Discussion} \label{sec:discussion}
So far, we have used a carefully controlled sample and quantitative statistical methods to investigate how properties of a galaxy group and its galaxy members depend on its structure parameter: group compactness.  However, there are still some biases remaining in the data analysis pipeline, which could potentially affect our results. In this section, we further discuss on these issues and present the robustness of our results.

\subsection{Redshift Incompleteness} \label{sec:contamination}
As mentioned in section~\ref{sec:compactness}, redshift incompleteness are more frequently occurred in the HC groups, figure~\ref{fig:zcomp} presents the redshift completeness as a function of $M_h$ for the HC, IC, and LC groups. A trend of increasing redshift incompleteness with decreasing $M_h$ is clearly seen for the HC groups, while there is no such trend for the IC and LC groups. This systematic trend is a result of increasing compactness limit for the selection of HC groups in low $M_h$ bins (figure~\ref{fig:compactbin}), where the fiber collision effect is easier to be trigger.  

In our study, we have assigned redshifts from their nearest neighbor for these galaxies without any spectroscopic redshifts. That is to say, there is a larger fraction of galaxies in the HC groups being contaminated by foreground or background galaxies, which biases the $M_h$ estimation of HC groups to higher values. This bias introduces a question on our results in section~\ref{sec:x-ray}, where we have shown that the HC groups have systematically fainter X-ray luminosities than LC groups at lower $M_h$ bins. In section~\ref{sec:data}, we show that $\sim 65 \%$ of the spectroscopy of unmeasured galaxies are genuine members of their assigned groups. Therefore, we do not expect that this redshift incompleteness bias on $M_h$ estimation will be significant. For the lowest $M_h$ bin with the most serious incompleteness problem, the redshift completeness of the HC groups is $\sim 78 \%$, which results in an over-estimation of $M_h$ for the HC groups at the level of $\sim 0.03$ dex, much smaller than the bin width ($\sim 0.63$ dex). Such a small difference in $M_h$ obviously can not account for the difference in $L_{\rm X}$ between the low-mass HC and LC groups ($\sim 0.3$ dex) we have obtained in section~\ref{sec:data}.

Besides the global bias, as shown by the subplots in the inset of figure~\ref{fig:zcomp}, the redshift incompleteness is not uniform, but clearly biased toward fainter (or satellite) galaxies. For instance, the redshift completeness of the central galaxies and satellites of HC groups in the lowest $M_h$ bin are $\sim 87 \%$ and $\sim 72 \%$, respectively.\footnote{This less significant bias might stem from the fact that the brighter galaxies more easily obtain spectroscopic reshifts from the LAMOST spectral survey easier \citep{Shen..2016}.} This non-uniform bias in magnitude might have an impact on the CLF and magnitude gap studies in section~\ref{sec:global}. 

To further validate our results presented in section~\ref{sec:global} compared with the redshift incompleteness, we use a smaller but cleaner subsample for the robustness test. To do that, we remove the groups with member completeness $\le 80 \%$ to ensure that low richness ($3 \le N \le 5$) groups have $100\%$ completeness and the bias on $M_h$ estimation for the high richness groups ($N \ge 6$) are negligible \footnote{Most of the groups with $M_h \lesssim 10^{13.5}h^{-1}M_{\odot}$ have richness $3 \le N \le 5$.}. After removing $\sim 5,500$ groups not satisfying with this condition, we repeat all the studies in section~\ref{sec:global} and find consistent results inside statistical errors.
 
Finally, during the study of the properties of group member galaxies in section~\ref{sec:properties}, we only use the galaxies with spectroscopic measurements in SDSS-DR7 and make comparisons using carefully selected control samples. Therefore, we do not expect any biases to be produced by the redshift incompleteness here.

\begin{figure}
  \centering
  \subfigure{
    \includegraphics[width=1.\columnwidth]{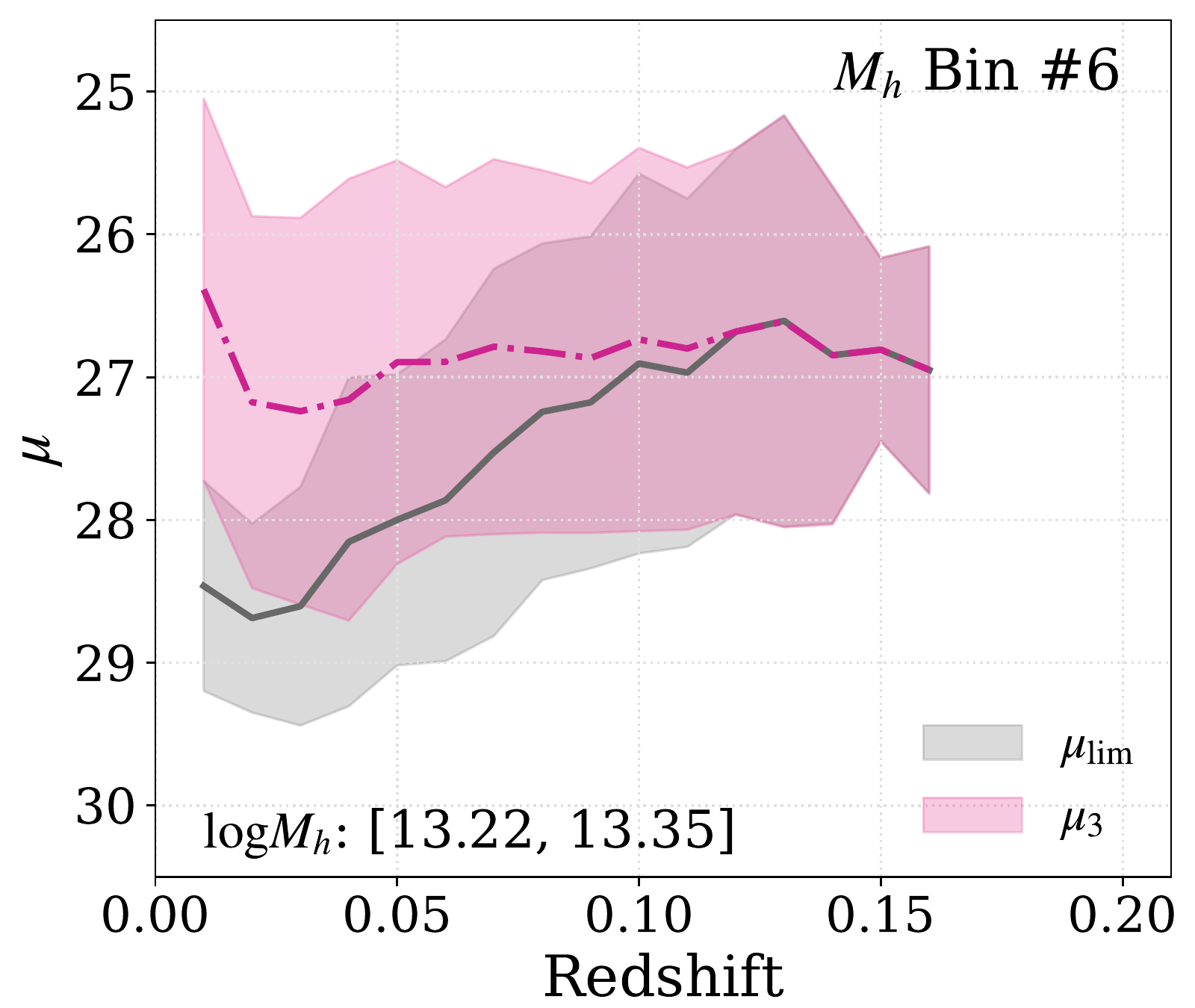}}
  \caption{An example of the median and scatter of $\mu_{\rm lim}$ (grey) and $\mu_3$ (magenta) as a function of redshift for Y07 groups in the $M_h$ Bin $= 6$.}
  \label{fig:mu345}
\end{figure}

\subsection{Alternative Compactness Definition: $\mu_{N}$}
Historically, \citet{Hickson1982} originally used a fixed value of $\mu_{\rm lim} = 26.0$ mag\,arcsec$^{-2}$ to define the classical compact groups. However, as mentioned in section~\ref{sec:compactness}, the $\mu_{\rm lim}$ distribution shows a clear dependence on redshift. This redshift bias might be due to the fact that galaxies within groups are not randomly distributed. The brighter galaxy members suffer the dynamical friction effect more significantly and are on average located in more central regions \citep{Saro..2013}. The faint members in the outskirts are more likely to be out of the detection limit in the flux limit sample, which makes the $\mu_{\rm lim}$ measurement be biased toward lower values for higher redshift groups. Therefore, using a fixed value of $\mu_{\rm lim}$ as the compact group threshold will result in an in-homogeneous sample at different redshifts \citep{Paper1}. One option that might alleviate this redshift bias is adopting an alternative compactness definition, $\mu_{N}$, based on the brightest $N$ galaxy members.

Here, we use $\mu_{3}$, the surface brightness as given by Equation (\ref{eq:mulim}) but calculated using the three brightest galaxies only, to test our hypothesis. We show an example of the $\mu_3-$redshift relation for groups in $M_h$ Bin$= 6$ in figure~\ref{fig:mu345}. Clearly, the redshift dependence is significantly reduced, and the quartile values show little change with redshift. On the other hand, using fewer galaxy members to calculate $\mu_N$ leads to a larger random dispersion on group compactness. As can be seen in figure~\ref{fig:mu345}, the scatter of $\mu_3$ is about 2 times larger than $\mu_{\rm lim}$ at low redshift. At high redshift, because of the decreasing of the observable members, $\mu_3$ becomes to be equivalent to $\mu_{\rm lim}$. We have tested, using $\mu_3$ instead of $\mu_{\rm lim}$ to define the HC and LC groups, and all of our conclusions remain unchanged but with much lower significance.

\subsection{Magnitude Measurement of Galaxies in the High Density Region} \label{sec:magbias}
The magnitude measurement of galaxies in the high density region is known to be easily biased to high values, i.e., underestimation of the flux because of the difficulties in the subtraction of background \citep[e.g,][]{Lauer..2007, Linden..2007}. 

For our HC groups, the surface brightness is as high as $\mu_{\rm lim}\sim 26$ mag arcsec$^{-2}$, where the magnitude measurement bias very likely happens. If so, the stellar masses of galaxy members of the HC groups, especially for the central population, will be underestimated. Such an underestimation effect might account for the systematically higher (although not significant) quenching fraction of central galaxies within higher $M_h$ HC groups. On the other hand, because of the underestimation of $M_{\star}$, $M_h$ will also be underestimated. Such an underestimation effect on $M_h$ might compensate for the overestimation effect discussed in section~\ref{sec:contamination}.

\section{Conclusion} \label{sec:conclusion}
In this study, we have performed a systematic and statistical analysis of the compactness of galaxy groups based on an $N \ge 3$ group sample drawn from the  halo-based group catalog of \citet{Yang..2007} as well as the extra GAMA and LAMOST redshift measurements. Based on the group surface brightness $\mu_{\rm lim}$, the total luminosity averaged over the smallest circle enclosing all the members, akin to the compactness criterion for classical compact groups introduced by \citet{Hickson1982}, we divide the samples into the HC and LC groups based on the aggregate relation of $\mu_{\rm lim}$ versus redshift in each $M_h$ bin. We investigate various aspects of the individual and global properties such as CLF, $\Delta m_{12}$, X-ray luminosity of ICM, quenched fraction and AGN fraction. We reach the following conclusions and predictions:

\begin{enumerate}
	\item For massive halos ($M_h \gtrsim 10^{13.5}h^{-1}M_{\odot}$),  there is no significant dependence on group compactness. Also, at the same redshift, massive groups are generally less compact (higher $\mu_{\rm lim}$) than small groups in observation. This echoes the fact that the classical compact groups are generally small galaxy systems \citep[e.g.,][]{McConnachie..2009, Paper1}.
	\item The HC groups are generally comprised of several members with comparable $M_{\star}$ (or luminosity), while the LC groups generally contain a more dominant central galaxy. This is relevant to the studies of fossil groups, such as the studies by \citet{Farhang..2017} and \citet{Raouf..2019}, which have suggested that the fossil groups with a larger magnitude gap $\Delta m_{12}$ tend to be older and the spatial distribution of their galaxy members are less crowded compared to the non-relaxed groups. Combined with the fact that the average X-ray luminosity of the LC groups is higher than the HC ones in the range of $M_h \lesssim 10^{13.5}h^{-1}M_{\odot}$, we argue that those small HC groups, on average, are in the early stages of group merging.
	\item The HC groups contain more passive galaxies than the LC groups with similar $M_h$ at a given $M_{\star}$, which is a result of their different halo-centric distance distributions for a satellite population. When controlled for halo-centric distance further, this difference tends to disappear. We conclude that the halo quenching effect in galaxy members, which makes the dependence of passive fraction on halo-centric distance, is a faster process compared to the dynamical relaxing time-scale for the groups in which they are hosted.
	\item The fraction of galaxies that host an optical or radio-loud AGN depends weakly on group compactness, indicating the local environmental effect triggered by group evolution does not correlate with the AGN activity of galaxies.
\end{enumerate}

In summary, our results imply a relationship between the observed group compactness and halo formation history. However,  this relationship is not consistent with the inferences obtained from simulations of pure dark matter halo, where the halos with higher concentration  are believed to be formed earlier \citep[e.g.,][]{ZhaoDH..2009, Ragagnin..2019}. To further clarify the possible inconsistency between our observations and numerical simulations, we need a joint effort for both. On the one hand, by expanding to a larger and deeper galaxy sample \citep[e.g., DESI, ][]{DESI}, we would expand the definition of group compactness to include more fainter group members. On the other hand, cosmological simulations including both dark matter and baryonic components \citep[e.g., IllustrisTNG, ][]{IllustrisTNG} are helpful for resolving the formation histories of groups with different group compactness and exploring the relationship between halo concentration and group compactness.

\section*{Acknowledgements}
We acknowledge the anonymous referee for helpful comments and detailed suggestions. This work is supported by National Key R\&D Program of China No.2019YFA0405501 and the National Natural Science Foundation of China (grant No. 12073059, 12103017).

This work has made use of data products from the Sloan Digital Sky Survey (SDSS, \url{http://www.sdss.org}), the Large Sky Area Multi-Object Fiber Spectroscopic Telescope (LAMOST, \url{http://www.lamost.org}), the GAMA survey (\url{http://www.gama-survey.org}. We are thankful for their tremendous efforts on the surveying work.

The Guoshoujing Telescope (the Large Sky Area Multi-Object Fiber Spectroscopic Telescope LAMOST) is a National Major Scientific Project built by the Chinese Academy of Sciences. Funding for the project has been provided by the National Development and Reform Commission. LAMOST is operated and managed by the National Astronomical Observatories, Chinese Academy of Sciences.

\end{CJK*}

\begin{thebibliography}{}
\expandafter\ifx\csname natexlab\endcsname\relax\def\natexlab#1{#1}\fi
\providecommand{\url}[1]{\href{#1}{#1}}
\providecommand{\dodoi}[1]{doi:~\href{http://doi.org/#1}{\nolinkurl{#1}}}
\providecommand{\doeprint}[1]{\href{http://ascl.net/#1}{\nolinkurl{http://ascl.net/#1}}}
\providecommand{\doarXiv}[1]{\href{https://arxiv.org/abs/#1}{\nolinkurl{https://arxiv.org/abs/#1}}}

\bibitem[Abazajian et al.(2009)]{SDSSDR7} Abazajian, K.~N., Adelman-McCarthy, J.~K., Ag{\"u}eros, M.~A., et al.\ 2009, \apjs, 182, 543. doi:10.1088/0067-0049/182/2/543
\bibitem[Ahumada et al.(2020)]{SDSSDR16} Ahumada, R., Prieto, C.~A., Almeida, A., et al.\ 2020, \apjs, 249, 3. doi:10.3847/1538-4365/ab929e
\bibitem[Amram et al.(2004)]{Amram..2004} Amram, P., Mendes de Oliveira, C., Plana, H., et al.\ 2004, \apjl, 612, L5. doi:10.1086/424482
\bibitem[Argudo-Fern{\'a}ndez et al.(2016)]{Argudo..2016} Argudo-Fern{\'a}ndez, M., Shen, S., Sabater, J., et al.\ 2016, \aap, 592, A30. doi:10.1051/0004-6361/201628232
\bibitem[Baldwin et al.(1981)]{BPT} Baldwin, J.~A., Phillips, M.~M., \& Terlevich, R.\ 1981, \pasp, 93, 5. doi:10.1086/130766
\bibitem[Barnes(1989)]{Barnes1989} Barnes, J.~E.\ 1989, \nat, 338, 123. doi:10.1038/338123a0
\bibitem[Barton et al.(1996)]{Barton..1996} Barton, E., Geller, M., Ramella, M., Marzke, R.~O., \& da Costa, L.~N.\ 1996, \aj, 112, 871 
\bibitem[Becker et al.(1995)]{Becker..1995} Becker, R.~H., White, R.~L., \& Helfand, D.~J.\ 1995, \apj, 450, 559. doi:10.1086/176166
\bibitem[Bell et al.(2003)]{Bell..2003} Bell, E.~F., McIntosh, D.~H., Katz, N., et al.\ 2003, \apjs, 149, 289. doi:10.1086/378847
\bibitem[Best et al.(2005)]{Best..2005} Best, P.~N., Kauffmann, G., Heckman, T.~M., et al.\ 2005, \mnras, 362, 25. doi:10.1111/j.1365-2966.2005.09192.x
\bibitem[Best \& Heckman(2012)]{Best.Heckman2012} Best, P.~N. \& Heckman, T.~M.\ 2012, \mnras, 421, 1569. doi:10.1111/j.1365-2966.2012.20414.x
\bibitem[Blanton et al.(2005)]{NYU-VAGC} Blanton, M.~R., Schlegel, D.~J., Strauss, M.~A., et al.\ 2005, \aj, 129, 2562 
\bibitem[Bluck et al.(2014)]{Bluck..2014} Bluck, A.~F.~L., Mendel, J.~T., Ellison, S.~L., et al.\ 2014, \mnras, 441, 599. doi:10.1093/mnras/stu594
\bibitem[Bluck et al.(2016)]{Bluck..2016} Bluck, A.~F.~L., Mendel, J.~T., Ellison, S.~L., et al.\ 2016, \mnras, 462, 2559. doi:10.1093/mnras/stw1665
\bibitem[Briel et al.(1996)]{ROSAT} Briel, U.~G., Aschenbach, B., Hasinger, G., et al.\ 1996, The ROSAT Users' Handbook. A guide to the Astronomical X-Ray satellite ROSAT. Intended to provide ROSAT users with an overview of all technical aspects of ROSAT. With special emphasis on the calibration of the various instruments. Garching bei Muenchen, 1996. 212 pages.
\bibitem[Brinchmann et al.(2004)]{MPA-JHU} Brinchmann, J., Charlot, S., White, S.~D.~M., et al.\ 2004, \mnras, 351, 1151. doi:10.1111/j.1365-2966.2004.07881.x
\bibitem[B{\"o}hringer et al.(2000)]{Bohringer..2000} B{\"o}hringer, H., Voges, W., Huchra, J.~P., et al.\ 2000, \apjs, 129, 435. doi:10.1086/313427
\bibitem[B{\"o}hringer et al.(2004)]{Bohringer..2004} B{\"o}hringer, H., Schuecker, P., Guzzo, L., et al.\ 2004, \aap, 425, 367. doi:10.1051/0004-6361:20034484
\bibitem[Charlot \& Fall(2000)]{Charlot.Fall2000} Charlot, S. \& Fall, S.~M.\ 2000, \apj, 539, 718. doi:10.1086/309250
\bibitem[Cid Fernandes et al.(2010)]{CF..2010} Cid Fernandes, R., Stasi{\'n}ska, G., Schlickmann, M.~S., et al.\ 2010, \mnras, 403, 1036. doi:10.1111/j.1365-2966.2009.16185.x
\bibitem[Cid Fernandes et al.(2011)]{CF..2011} Cid Fernandes, R., Stasi{\'n}ska, G., Mateus, A., et al.\ 2011, \mnras, 413, 1687. doi:10.1111/j.1365-2966.2011.18244.x
\bibitem[Cluver et al.(2020)]{Cluver..2020} Cluver, M.~E., Jarrett, T.~H., Taylor, E.~N., et al.\ 2020, arXiv:2006.07535
\bibitem[Coenda et al.(2012)]{Coenda..2012} Coenda, V., Muriel, H., \& Mart{\'{\i}}nez, H.~J.\ 2012, \aap, 543, A119 \bibitem[Coenda et al.(2015)]{Coenda..2015} Coenda, V., Muriel, H., \& Mart{\'\i}nez, H.~J.\ 2015, \aap, 573, A96. doi:10.1051/0004-6361/201424870
\bibitem[Condon et al.(1998)]{Condon..1998} Condon, J.~J., Cotton, W.~D., Greisen, E.~W., et al.\ 1998, \aj, 115, 1693. doi:10.1086/300337
\bibitem[Cooray(2006)]{Cooray2006} Cooray, A.\ 2006, \mnras, 365, 842. doi:10.1111/j.1365-2966.2005.09747.x
\bibitem[Coziol \& Plauchu-Frayn(2007)]{Coziol.Plauchu-Frayn2007} Coziol, R. \& Plauchu-Frayn, I.\ 2007, \aj, 133, 2630. doi:10.1086/513514
\bibitem[Croton et al.(2006)]{Croton..2006} Croton, D.~J., Springel, V., White, S.~D.~M., et al.\ 2006, \mnras, 365, 11. doi:10.1111/j.1365-2966.2005.09675.x
\bibitem[Cui et al.(2021)]{CuiWG..2021} Cui, W., Dav{\'e}, R., Peacock, J.~A., et al.\ 2021, Nature Astronomy. doi:10.1038/s41550-021-01404-1
\bibitem[Dai et al.(2007)]{DaiXY..2007} Dai, X., Kochanek, C.~S., \& Morgan, N.~D.\ 2007, \apj, 658, 917. doi:10.1086/509651
\bibitem[Dariush et al.(2010)]{Dariush..2010} Dariush, A.~A., Raychaudhury, S., Ponman, T.~J., et al.\ 2010, \mnras, 405, 1873. doi:10.1111/j.1365-2966.2010.16569.x
\bibitem[Dark Energy Survey Collaboration et al.(2016)]{DESI} Dark Energy Survey Collaboration, Abbott, T., Abdalla, F.~B., et al.\ 2016, \mnras, 460, 1270. doi:10.1093/mnras/stw641
\bibitem[Davies et al.(2019)]{Davies..2019} Davies, L.~J.~M., Robotham, A.~S.~G., Lagos, C. del P., et al.\ 2019, \mnras, 483, 5444. doi:10.1093/mnras/sty3393
\bibitem[Dekel \& Birnboim(2006)]{Dekel.Birnboim2006} Dekel, A. \& Birnboim, Y.\ 2006, \mnras, 368, 2. doi:10.1111/j.1365-2966.2006.10145.x
\bibitem[D{\'\i}az-Gim{\'e}nez \& Zandivarez(2015)]{DiazGimenez.Zandivarez2015} D{\'\i}az-Gim{\'e}nez, E. \& Zandivarez, A.\ 2015, \aap, 578, A61. doi:10.1051/0004-6361/201425267
\bibitem[D{\'\i}az-Gim{\'e}nez et al.(2018)]{DiazGimenez..2018} D{\'\i}az-Gim{\'e}nez, E., Zandivarez, A., \& Taverna, A.\ 2018, \aap, 618, A157. doi:10.1051/0004-6361/201833329
\bibitem[Dickey \& Lockman(1990)]{Dickey.Lockman1990} Dickey, J.~M. \& Lockman, F.~J.\ 1990, \araa, 28, 215. doi:10.1146/annurev.aa.28.090190.001243
\bibitem[Eckert et al.(2011)]{Eckert..2011} Eckert, D., Molendi, S., \& Paltani, S.\ 2011, \aap, 526, A79. doi:10.1051/0004-6361/201015856
\bibitem[Efstathiou et al.(1988)]{Efstathiou..1988} Efstathiou, G., Ellis, R.~S., \& Peterson, B.~A.\ 1988, \mnras, 232, 431
\bibitem[Fabian(2012)]{Fabian2012} Fabian, A.~C.\ 2012, \araa, 50, 455. doi:10.1146/annurev-astro-081811-125521
\bibitem[Farhang et al.(2017)]{Farhang..2017} Farhang, A., Khosroshahi, H.~G., Mamon, G.~A., et al.\ 2017, \apj, 840, 58. doi:10.3847/1538-4357/aa6b00
\bibitem[Farouki \& Shapiro(1981)]{Farouki.Shapiro1981} Farouki, R., \& Shapiro, S.~L.\ 1981, \apj, 243, 32
\bibitem[Feng et al.(2019)]{Feng..2019} Feng, S., Shen, S.-Y., Yuan, F.-T., et al.\ 2019, \apj, 880, 114
\bibitem[Gunn \& Gott(1972)]{Gunn.Gott1972} Gunn, J.~E., \& Gott, J.~R.\ 1972, \apj, 176, 1
\bibitem[Guo et al.(2010)]{GuoQ..2010} Guo, Q., White, S., Li, C., et al.\ 2010, \mnras, 404, 1111. doi:10.1111/j.1365-2966.2010.16341.x
\bibitem[Haas et al.(2012)]{Haas..2012} Haas, M.~R., Schaye, J., \& Jeeson-Daniel, A.\ 2012, \mnras, 419, 2133. doi:10.1111/j.1365-2966.2011.19863.x
\bibitem[Harrison(2017)]{Harrison2017} Harrison, C.~M.\ 2017, Nature Astronomy, 1, 0165. doi:10.1038/s41550-017-0165
\bibitem[Henriques et al.(2017)]{Henriques..2017} Henriques, B.~M.~B., White, S.~D.~M., Thomas, P.~A., et al.\ 2017, \mnras, 469, 2626. doi:10.1093/mnras/stx1010
\bibitem[Hickson(1982)]{Hickson1982} Hickson, P.\ 1982, \apj, 255, 382 
\bibitem[Jiang et al.(2014)]{JiangCY..2014} Jiang, C.~Y., Jing, Y.~P., \& Han, J.\ 2014, \apj, 790, 7. doi:10.1088/0004-637X/790/1/7
\bibitem[Jing et al.(1998)]{JingYP..1998} Jing, Y.~P., Mo, H.~J., \& B{\"o}rner, G.\ 1998, \apj, 494, 1. doi:10.1086/305209
\bibitem[Kauffmann et al.(2003)]{MPA-JHU-K} Kauffmann, G., Heckman, T.~M., White, S.~D.~M., et al.\ 2003, \mnras, 341, 33. doi:10.1046/j.1365-8711.2003.06291.x
\bibitem[Kauffmann et al.(2003)]{K03} Kauffmann, G., Heckman, T.~M., Tremonti, C., et al.\ 2003, \mnras, 346, 1055. doi:10.1111/j.1365-2966.2003.07154.x
\bibitem[Kauffmann et al.(2004)]{Kauffmann..2004} Kauffmann, G., White, S.~D.~M., Heckman, T.~M., et al.\ 2004, \mnras, 353, 713. doi:10.1111/j.1365-2966.2004.08117.x
\bibitem[Kauffmann et al.(2013)]{Kauffmann..2013} Kauffmann, G., Li, C., Zhang, W., et al.\ 2013, \mnras, 430, 1447. doi:10.1093/mnras/stt007
\bibitem[Kewley et al.(2001)]{K01} Kewley, L.~J., Dopita, M.~A., Sutherland, R.~S., et al.\ 2001, \apj, 556, 121. doi:10.1086/321545
\bibitem[Khosroshahi et al.(2017)]{Khosroshahi..2017} Khosroshahi, H.~G., Raouf, M., Miraghaei, H., et al.\ 2017, \apj, 842, 81. doi:10.3847/1538-4357/aa7048
\bibitem[Knobel et al.(2015)]{Knobel..2015} Knobel, C., Lilly, S.~J., Woo, J., et al.\ 2015, \apj, 800, 24. doi:10.1088/0004-637X/800/1/24
\bibitem[Koester et al.(2007)]{Koester..2007} Koester, B.~P., McKay, T.~A., Annis, J., et al.\ 2007, \apj, 660, 239. doi:10.1086/509599
\bibitem[Kravtsov et al.(2004)]{Kravtsov..2004} Kravtsov, A.~V., Berlind, A.~A., Wechsler, R.~H., et al.\ 2004, \apj, 609, 35. doi:10.1086/420959
\bibitem[Lange et al.(2018)]{Lange..2018} Lange, J.~U., van den Bosch, F.~C., Hearin, A., et al.\ 2018, \mnras, 473, 2830. doi:10.1093/mnras/stx2434
\bibitem[Lauer et al.(2007)]{Lauer..2007} Lauer, T.~R., Faber, S.~M., Richstone, D., et al.\ 2007, \apj, 662, 808. doi:10.1086/518223
\bibitem[Li et al.(2006)]{LiCheng..2006} Li, C., Kauffmann, G., Jing, Y.~P., et al.\ 2006, \mnras, 368, 21. doi:10.1111/j.1365-2966.2006.10066.x
\bibitem[Li et al.(2020)]{LiPF..2020} Li, P., Wang, H., Mo, H.~J., et al.\ 2020, \apj, 902, 75. doi:10.3847/1538-4357/abb66c
\bibitem[Li \& Cen(2020)]{Li.Cen2020} Li, Z. \& Cen, R.\ 2020, \apj, 898, 39. doi:10.3847/1538-4357/ab9811
\bibitem[Liske et al.(2015)]{gama} Liske, J., Baldry, I.~K., Driver, S.~P., et al.\ 2015, \mnras, 452, 2087
\bibitem[Liu et al.(2010)]{LiuL..2010} Liu, L., Yang, X., Mo, H.~J., et al.\ 2010, \apj, 712, 734. doi:10.1088/0004-637X/712/1/734
\bibitem[Liu et al.(2019)]{LiuCX..2019} Liu, C., Hao, L., Wang, H., et al.\ 2019, \apj, 878, 69. doi:10.3847/1538-4357/ab1ea0
\bibitem[Lovisari et al.(2021)]{Lovisari..2021} Lovisari, L., Ettori, S., Gaspari, M., et al.\ 2021, Universe, 7, 139. doi:10.3390/universe7050139
\bibitem[Lu et al.(2014)]{LuZK..2014} Lu, Z., Mo, H.~J., Lu, Y., et al.\ 2014, \mnras, 439, 1294. doi:10.1093/mnras/stu016
\bibitem[Luo et al.(2015)]{lamost} Luo, A.-L., Zhao, Y.-H., Zhao, G., et al.\ 2015, Research in Astronomy and Astrophysics, 15, 1095 
\bibitem[Mamon(1993)]{Mamon1993} Mamon, G.\ 1993, N-body Problems and Gravitational Dynamics, 188
\bibitem[Mamon(2007)]{Mamon2007} Mamon, G.~A.\ 2007, Groups of Galaxies in the Nearby Universe, 203
\bibitem[Mandelbaum et al.(2016)]{Mandelbaum..2016} Mandelbaum, R., Wang, W., Zu, Y., et al.\ 2016, \mnras, 457, 3200. doi:10.1093/mnras/stw188
\bibitem[Mantz et al.(2010)]{Mantz..2010} Mantz, A., Allen, S.~W., Ebeling, H., et al.\ 2010, \mnras, 406, 1773. doi:10.1111/j.1365-2966.2010.16993.x
\bibitem[Mart{\'\i}nez \& Zandivarez(2012)]{Martinez.Zandivarez2012} Mart{\'\i}nez, H.~J. \& Zandivarez, A.\ 2012, \mnras, 419, L24. doi:10.1111/j.1745-3933.2011.01170.x
\bibitem[Matthee et al.(2017)]{Matthee..2017} Matthee, J., Schaye, J., Crain, R.~A., et al.\ 2017, \mnras, 465, 2381. doi:10.1093/mnras/stw2884
\bibitem[McConnachie et al.(2009)]{McConnachie..2009} McConnachie, A.~W., Patton, D.~R., Ellison, S.~L., \& Simard, L.\ 2009, \mnras, 395, 255 
\bibitem[Mendel et al.(2011)]{Mendel..2011} Mendel, J.~T., Ellison, S.~L., Simard, L., Patton, D.~R., \& McConnachie, A.~W.\ 2011, \mnras, 418, 1409 
\bibitem[Mo et al.(1999)]{MoHJ..1999} Mo, H.~J., Mao, S., \& White, S.~D.~M.\ 1999, \mnras, 304, 175. doi:10.1046/j.1365-8711.1999.02289.x
\bibitem[Moore et al.(1996)]{Moore..1996} Moore, B., Katz, N., Lake, G., Dressler, A., \& Oemler, A.\ 1996, \nat, 379, 613 
\bibitem[More et al.(2011)]{More..2011} More, S., van den Bosch, F.~C., Cacciato, M., et al.\ 2011, \mnras, 410, 210. doi:10.1111/j.1365-2966.2010.17436.x
\bibitem[More(2012)]{More..2012} More, S.\ 2012, \apj, 761, 127. doi:10.1088/0004-637X/761/2/127
\bibitem[Moster et al.(2018)]{Moster..2018} Moster, B.~P., Naab, T., \& White, S.~D.~M.\ 2018, \mnras, 477, 1822. doi:10.1093/mnras/sty655
\bibitem[Osmond \& Ponman(2004)]{Osmond.Ponman2004} Osmond, J.~P.~F. \& Ponman, T.~J.\ 2004, \mnras, 350, 1511. doi:10.1111/j.1365-2966.2004.07742.x
\bibitem[Osterbrock(1989)]{Osterbrock1989} Osterbrock, D.~E.\ 1989, Astrophysics of Gaseous Nebulae and Active Galactic Nuclei, by Donald E. Osterbrock. Published by University Science Books, ISBN 0-935702-22-9, 408pp, 1989.
\bibitem[Paranjape \& Sheth(2012)]{Paranjape.Sheth2012} Paranjape, A. \& Sheth, R.~K.\ 2012, \mnras, 423, 1845. doi:10.1111/j.1365-2966.2012.21008.x
\bibitem[Pasquali et al.(2009)]{Pasquali..2009} Pasquali, A., van den Bosch, F.~C., Mo, H.~J., et al.\ 2009, \mnras, 394, 38. doi:10.1111/j.1365-2966.2008.14233.x
\bibitem[Peacock \& Smith(2000)]{Peacock.Smith2000} Peacock, J.~A. \& Smith, R.~E.\ 2000, \mnras, 318, 1144. doi:10.1046/j.1365-8711.2000.03779.x
\bibitem[Peng et al.(2010)]{PengYJ..2010} Peng, Y.-. jie ., Lilly, S.~J., Kova{\v{c}}, K., et al.\ 2010, \apj, 721, 193. doi:10.1088/0004-637X/721/1/193
\bibitem[Peng et al.(2012)]{PengYJ..2012} Peng, Y.-. jie ., Lilly, S.~J., Renzini, A., et al.\ 2012, \apj, 757, 4. doi:10.1088/0004-637X/757/1/4
\bibitem[Peng et al.(2014)]{PengYJ..2014} Peng, Y.-. jie ., Lilly, S.~J., Renzini, A., et al.\ 2014, \apj, 790, 95. doi:10.1088/0004-637X/790/2/95
\bibitem[Pillepich et al.(2018)]{IllustrisTNG} Pillepich, A., Springel, V., Nelson, D., et al.\ 2018, \mnras, 473, 4077. doi:10.1093/mnras/stx2656
\bibitem[Ragagnin et al.(2019)]{Ragagnin..2019} Ragagnin, A., Dolag, K., Moscardini, L., et al.\ 2019, \mnras, 486, 4001. doi:10.1093/mnras/stz1103
\bibitem[Raouf et al.(2014)]{Raouf..2014} Raouf, M., Khosroshahi, H.~G., Ponman, T.~J., et al.\ 2014, \mnras, 442, 1578. doi:10.1093/mnras/stu963
\bibitem[Raouf et al.(2018)]{Raouf..2018} Raouf, M., Khosroshahi, H.~G., Mamon, G.~A., et al.\ 2018, \apj, 863, 40. doi:10.3847/1538-4357/aace57
\bibitem[Raouf et al.(2019)]{Raouf..2019} Raouf, M., Smith, R., Khosroshahi, H.~G., et al.\ 2019, \apj, 887, 264. doi:10.3847/1538-4357/ab5581
\bibitem[Reddick et al.(2013)]{Reddick..2013} Reddick, R.~M., Wechsler, R.~H., Tinker, J.~L., et al.\ 2013, \apj, 771, 30. doi:10.1088/0004-637X/771/1/30
\bibitem[Robotham et al.(2011)]{Robotham..2011} Robotham, A.~S.~G., Norberg, P., Driver, S.~P., et al.\ 2011, \mnras, 416, 2640. doi:10.1111/j.1365-2966.2011.19217.x
\bibitem[Saro et al.(2013)]{Saro..2013} Saro, A., Mohr, J.~J., Bazin, G., et al.\ 2013, \apj, 772, 47. doi:10.1088/0004-637X/772/1/47
\bibitem[Schechter(1976)]{Schechter1976} Schechter, P.\ 1976, \apj, 203, 297
\bibitem[Shen et al.(2008)]{Shen..2008} Shen, S., Kauffmann, G., von der Linden, A., et al.\ 2008, \mnras, 389, 1074. doi:10.1111/j.1365-2966.2008.13647.x
\bibitem[Shen et al.(2014)]{Shen..2014} Shen, S., Yang, X., Mo, H., et al.\ 2014, \apj, 782, 23. doi:10.1088/0004-637X/782/1/23
\bibitem[Shen et al.(2016)]{Shen..2016} Shen, S.-Y., Argudo-Fern{\'a}ndez, M., Chen, L., et al.\ 2016, Research in Astronomy and Astrophysics, 16, 43
\bibitem[Simha et al.(2012)]{Simha..2012} Simha, V., Weinberg, D.~H., Dav{\'e}, R., et al.\ 2012, \mnras, 423, 3458. doi:10.1111/j.1365-2966.2012.21142.x
\bibitem[Sohn et al.(2015)]{Sohn..2015} Sohn, J., Hwang, H.~S., Geller, M.~J., et al.\ 2015, Journal of Korean Astronomical Society, 48, 381 
\bibitem[Sohn et al.(2016)]{Sohn..2016} Sohn, J., Geller, M.~J., Hwang, H.~S., Zahid, H.~J., \& Lee, M.~G.\ 2016, \apjs, 225, 23 
\bibitem[Stewart et al.(2009)]{Stewart..2009} Stewart, K.~R., Bullock, J.~S., Barton, E.~J., et al.\ 2009, \apj, 702, 1005. doi:10.1088/0004-637X/702/2/1005
\bibitem[Tal et al.(2012)]{Tal..2012} Tal, T., Wake, D.~A., van Dokkum, P.~G., et al.\ 2012, \apj, 746, 138. doi:10.1088/0004-637X/746/2/138
\bibitem[Tammann et al.(1979)]{Tammann..1979} Tammann, G.~A., Yahil, A., \& Sandage, A.\ 1979, \apj, 234, 775
\bibitem[Torres-Flores et al.(2009)]{Torres-Flores..2009} Torres-Flores, S., Mendes de Oliveira, C., de Mello, D.~F., et al.\ 2009, \aap, 507, 723. doi:10.1051/0004-6361/200911878
\bibitem[Torres-Flores et al.(2010)]{Torres-Flores..2010} Torres-Flores, S., Mendes de Oliveira, C., Amram, P., et al.\ 2010, \aap, 521, A59. doi:10.1051/0004-6361/200913912
\bibitem[Torres-Flores et al.(2014)]{Torres-Flores..2014} Torres-Flores, S., Amram, P., Mendes de Oliveira, C., et al.\ 2014, \mnras, 442, 2188. doi:10.1093/mnras/stu1002
\bibitem[Trevisan \& Mamon(2017)]{Trevisan.Mamon2017} Trevisan, M. \& Mamon, G.~A.\ 2017, \mnras, 471, 2022. doi:10.1093/mnras/stx1656
\bibitem[Vale \& Ostriker(2006)]{Vale.Ostriker2006} Vale, A. \& Ostriker, J.~P.\ 2006, \mnras, 371, 1173. doi:10.1111/j.1365-2966.2006.10605.x
\bibitem[van den Bosch et al.(2007)]{vandenBosch..2007} van den Bosch, F.~C., Yang, X., Mo, H.~J., et al.\ 2007, \mnras, 376, 841. doi:10.1111/j.1365-2966.2007.11493.x
\bibitem[van den Bosch et al.(2008)]{vandenBosch..2008} van den Bosch, F.~C., Pasquali, A., Yang, X., et al.\ 2008, arXiv:0805.0002
\bibitem[Voges et al.(1999)]{Voges..1999} Voges, W., Aschenbach, B., Boller, T., et al.\ 1999, \aap, 349, 389
\bibitem[Von Der Linden et al.(2007)]{Linden..2007} Von Der Linden, A., Best, P.~N., Kauffmann, G., et al.\ 2007, \mnras, 379, 867. doi:10.1111/j.1365-2966.2007.11940.x
\bibitem[Wang et al.(2014)]{WangLei..2014} Wang, L., Yang, X., Shen, S., et al.\ 2014, \mnras, 439, 611. doi:10.1093/mnras/stt2481
\bibitem[Wang et al.(2018)]{WangEnci..2018} Wang, E., Wang, H., Mo, H., et al.\ 2018, \apj, 860, 102. doi:10.3847/1538-4357/aac4a5
\bibitem[Weinmann et al.(2006)]{Weinmann..2006} Weinmann, S.~M., van den Bosch, F.~C., Yang, X., et al.\ 2006, \mnras, 366, 2. doi:10.1111/j.1365-2966.2005.09865.x
\bibitem[Wen \& Han(2013)]{Wen.Han2013} Wen, Z.~L. \& Han, J.~L.\ 2013, \mnras, 436, 275. doi:10.1093/mnras/stt1581
\bibitem[Wetzel et al.(2012)]{Wetzel..2012} Wetzel, A.~R., Tinker, J.~L., \& Conroy, C.\ 2012, \mnras, 424, 232. doi:10.1111/j.1365-2966.2012.21188.x
\bibitem[White et al.(1997)]{White..1997} White, D.~A., Jones, C., \& Forman, W.\ 1997, \mnras, 292, 419. doi:10.1093/mnras/292.2.419
\bibitem[Woo et al.(2013)]{Woo..2013} Woo, J., Dekel, A., Faber, S.~M., et al.\ 2013, \mnras, 428, 3306. doi:10.1093/mnras/sts274
\bibitem[Woo et al.(2015)]{Woo..2015} Woo, J., Dekel, A., Faber, S.~M., et al.\ 2015, \mnras, 448, 237. doi:10.1093/mnras/stu2755
\bibitem[Yang et al.(2003)]{Yang..2003} Yang, X., Mo, H.~J., \& van den Bosch, F.~C.\ 2003, \mnras, 339, 1057. doi:10.1046/j.1365-8711.2003.06254.x
\bibitem[Yang et al.(2005)]{Yang..2005} Yang, X., Mo, H.~J., van den Bosch, F.~C., et al.\ 2005, \mnras, 356, 1293
\bibitem[Yang et al.(2007)]{Yang..2007} Yang, X., Mo, H.~J., van den Bosch, F.~C., et al.\ 2007, \apj, 671, 153 
\bibitem[Yang et al.(2008)]{Yang..2008} Yang, X., Mo, H.~J., \& van den Bosch, F.~C.\ 2008, \apj, 676, 248. doi:10.1086/528954
\bibitem[Yang et al.(2012)]{Yang..2012} Yang, X., Mo, H.~J., van den Bosch, F.~C., et al.\ 2012, \apj, 752, 41
\bibitem[Yang et al.(2013)]{Yang..2013} Yang, X., Mo, H.~J., van den Bosch, F.~C., et al.\ 2013, \apj, 770, 115. doi:10.1088/0004-637X/770/2/115
\bibitem[Yang et al.(2021)]{Yang..2021} Yang, X., Xu, H., He, M., et al.\ 2021, \apj, 909, 143. doi:10.3847/1538-4357/abddb2
\bibitem[Zehavi et al.(2018)]{Zehavi..2018} Zehavi, I., Contreras, S., Padilla, N., et al.\ 2018, \apj, 853, 84. doi:10.3847/1538-4357/aaa54a
\bibitem[Zhang et al.(2021)]{ZhangCP..2021} Zhang, C., Peng, Y., Ho, L.~C., et al.\ 2021, \apj, 911, 57. doi:10.3847/1538-4357/abd723
\bibitem[Zhang et al.(2021)]{ZhangZW..2021} Zhang, Z., Wang, H., Luo, W., et al.\ 2021, \aap, 650, A155. doi:10.1051/0004-6361/202040150
\bibitem[Zhao et al.(2009)]{ZhaoDH..2009} Zhao, D.~H., Jing, Y.~P., Mo, H.~J., et al.\ 2009, \apj, 707, 354. doi:10.1088/0004-637X/707/1/354
\bibitem[Zheng \& Shen(2020)]{Paper1} Zheng, Y.-L., \& Shen, S.-Y.\ 2020, \apjs, 246, 12
\bibitem[Zheng \& Shen(2021)]{Paper2} Zheng, Y.-L. \& Shen, S.-Y.\ 2021, \apj, 911, 105. doi:10.3847/1538-4357/abeaa2
\bibitem[Zheng et al.(2005)]{ZhengZ..2005} Zheng, Z., Berlind, A.~A., Weinberg, D.~H., et al.\ 2005, \apj, 633, 791. doi:10.1086/466510
\bibitem[Zu \& Mandelbaum(2016)]{Zu.Mandelbaum2016} Zu, Y. \& Mandelbaum, R.\ 2016, \mnras, 457, 4360. doi:10.1093/mnras/stw221

\end{thebibliography}
\end{document}